\documentclass[journal]{IEEEtran}

\usepackage[swedish,english]{babel}
\usepackage{microtype}
\usepackage{cite}

\usepackage{graphicx,amsmath,url,amssymb,mathtools,steinmetz,bm}
\usepackage[hidelinks]{hyperref} 
\usepackage[nameinlink,capitalise]{cleveref}
\crefname{appsec}{Appendix}{Appendices}
\crefname{equation}{}{}
\usepackage{autonum} % \retainlabel{eqn:two} to keep label

\usepackage{bbold}
\usepackage[caption=false,font=footnotesize]{subfig}
\usepackage{siunitx}
\DeclareSIUnit\perunit{p.u.}
\DeclareSIUnit\voltampere{VA}
\DeclareSIUnit\wattsecond{Ws}
\DeclareSIUnit\watthour{Wh}

\usepackage{accents}

\usepackage[usenames, dvipsnames]{color}

\newtheorem{rem}{Remark}
\newtheorem{thm}{Theorem}

\newtheorem{prop}{Proposition}
\newtheorem{cor}{Corollary}

\newcommand{\T}{\textit{\textsf{T}}}
\newcommand{\WT}{WT}

\newcommand{\inv}{\ensuremath{{-1}}}

\newcommand{\complex}{\ensuremath{\mathbb{C}}}
\DeclareMathOperator{\Real}{\ensuremath{\mathrm{Re}}}

\newcommand{\jimag}{\ensuremath{j}}
\newcommand{\jomega}{\ensuremath{\jimag \omega}}

\newcommand{\s}{\ensuremath{(s)}}

\newcommand{\coi}{\ensuremath{\mathrm{COI}}}

\newcommand{\Lcoides}{\ensuremath{L}_\textrm{des}}

\usepackage{tikz}

\usepackage{scalerel}
\usepackage{tikz}
\usetikzlibrary{svg.path}

\definecolor{orcidlogocol}{HTML}{A6CE39}
\tikzset{
	orcidlogo/.pic={
		\fill[orcidlogocol] svg{M256,128c0,70.7-57.3,128-128,128C57.3,256,0,198.7,0,128C0,57.3,57.3,0,128,0C198.7,0,256,57.3,256,128z};
		\fill[white] svg{M86.3,186.2H70.9V79.1h15.4v48.4V186.2z}
		svg{M108.9,79.1h41.6c39.6,0,57,28.3,57,53.6c0,27.5-21.5,53.6-56.8,53.6h-41.8V79.1z M124.3,172.4h24.5c34.9,0,42.9-26.5,42.9-39.7c0-21.5-13.7-39.7-43.7-39.7h-23.7V172.4z}
		svg{M88.7,56.8c0,5.5-4.5,10.1-10.1,10.1c-5.6,0-10.1-4.6-10.1-10.1c0-5.6,4.5-10.1,10.1-10.1C84.2,46.7,88.7,51.3,88.7,56.8z};
	}
}

\newcommand\orcidicon[1]{\href{https://orcid.org/#1}{\mbox{\scalerel*{
				\begin{tikzpicture}[yscale=-1,transform shape]
				\pic{orcidlogo};
				\end{tikzpicture}
			}{|}}}}

\begin{document}

\frenchspacing

\title{
Dynamic Virtual Power Plant Design for Fast Frequency Reserves: Coordinating Hydro and Wind
}

\author{
	Joakim~Bj\"ork${\textsuperscript{\orcidicon{0000-0003-0656-7991}}}$,~\IEEEmembership{Student~Member,~IEEE,}
	Karl~Henrik~Johansson${\textsuperscript{\orcidicon{0000-0001-9940-5929}}}$,~\IEEEmembership{Fellow,~IEEE,}
	and
	Florian~D\"orfler${\textsuperscript{\orcidicon{0000-0002-9649-5305}}}$,~\IEEEmembership{Senior~Member,~IEEE}%<-this % stops a space
	
 \thanks{This work has been submitted to the IEEE for possible publication. Copyright may be transferred without notice, after which this version may no longer be accessible}%
\thanks{This work was supported by the KTH PhD program in the digitalization of electric power engineering and in part by the Knut and Alice
	Wallenberg Foundation, the Swedish Research Council, the Swedish Foundation for Strategic Research, and the European Union’s Horizon 2020 research and innovation programme under grant agreement No 883985.
}%
	\thanks{J. Bj\"ork and K. H. Johansson are with the School of Electrical Engineering and Computer Science, KTH Royal Institute of Technology, \mbox{100 44 Stockholm,} Sweden (email: joakbj@kth.se; kallej@kth.se).}% <-this % stops
	\thanks{F.~D\"orfler is with the Department of Information Technology and
Electrical Engineering, ETH Z\"urich, 8092 Z\"urich, Switzerland (e-mail:
dorfler@ethz.ch).}}

\maketitle

\selectlanguage{english}

%%%%%%%%%%%%%%%%%%%%%%%%%%%%%%%%%%%%%%%%%%%%%%%%%%%%%%%%%%%%

\begin{abstract}
To ensure frequency stability in future low-inertia power grids, fast ancillary services such as fast frequency reserves (FFR) have been proposed. In this work, the coordination of conventional (slow) frequency containment reserves (FCR) with FFR is treated as a decentralized model matching problem. The design results in a dynamic virtual power plant (DVPP) whose aggregated output fulfills the system operator (SO) requirements in all time scales, while accounting for the capacity and bandwidth limitation of participating devices. This is illustrated in a 5-machine representation of the Nordic synchronous grid. In the Nordic grid, stability issues and bandwidth limitations associated with non-minimum phase zeros of hydropower is a well-known problem. By simulating the disconnection of a \SI{1400}{\mega\watt} {importing dc link}, it is shown that the proposed DVPP design allows for coordinating fast FFR from wind, with slow FCR from hydro, while respecting dynamic limitations of all participating devices. The SO requirements are fulfilled in a realistic low-inertia scenario without the need to install battery storage or to waste wind energy by curtailing the wind turbines.
% presents a smooth response avoiding overshoot and secondary frequency dips during frequency recovery.}
\end{abstract}


\section{Introduction}
\IEEEPARstart{D}{eregulation} of the market and the transition towards renewable energy, is diversifying the mechanics behind electricity production. {Regulatory services provided by distributed energy resources coordinated as virtual plants are expected to be an important supplement to the services provided by large-scale power plants \cite{dall'aneseOptimalRegulationVirtual2018}.} At the same time, the frequency stability of grids are becoming more sensitive to load imbalances due to the growing share of converter-interfaced generation \cite{milanoFoundationsChallengesLowinertia2018}. 
A number of relatively recent blackouts are related to large frequency disturbances. The incidence of this phenomenon is expected to increase in the future as the energy transition continues; in fact they have doubled from the early 2000s \cite{rahmanLargestBlackoutsWorld2016}.
With growing shares of renewables, system operators (SOs) are therefore increasingly demanding renewable generation and other small-scale producers to participate in frequency containment reserves (FCR) \cite{brundlingerReviewAssessmentLatest2016}.

Virtual power plants (VPPs), aggregating together groups of small-scale producers and consumers, is a proposed solution to allow smaller players with more variable production to enter into the market with the functionality of a larger conventional power plant \cite{dall'aneseOptimalRegulationVirtual2018,sabooriVirtualPowerPlant2011,ghavidelReviewVirtualPower2016}. The main objectives are to coordinate dispatch, maximize the revenue, and to reduce the financial risk of variable generation, in the day-ahead and intra-day markets \cite{vasiraniAgentBasedApproachVirtual2013,alvarezGenericStorageModel2019}. But also other services, such as voltage regulation \cite{moutisVoltageRegulationSupport2018} and allocation of FCR resources \cite{alhelouPrimaryFrequencyResponse2020,schifferSynchronizationDroopcontrolledMicrogrids2013,zhongCoordinatedControlVirtual2021} have been proposed.

In this work, we design controllers that coordinate FCR over all time scales, beyond mere set-point tracking, forming a dynamic virtual power plant (DVPP) offering dynamic ancillary services \cite{posytyfConceptObjectives2021_manual}.
While none of the individual devices may be able to provide FCR consistently across all power and energy levels or over all time scales, a sufficiently heterogeneous ensemble will be able to do so.
Examples of heterogeneous devices complementing
each other while providing fast frequency reserves (FFR) include 
hydropower with initially inverse
response dynamics compensated by battery sources on short time scales
\cite{melkiInvestigationFrequencyContainment2019},
hybrid storage
pairing batteries with supercapacitors providing regulation on different time scales \cite{liPowerElectronicInterface2008,glavinStandalonePhotovoltaicSupercapacitor2008}, demand response \cite{mullerLargescaleDemonstrationPrecise2019}, or wind turbines ({\WT}s) \cite{morrenWindTurbinesEmulating2006,zhaoFastFrequencySupport2020} that can provide a quick response
but are subject to a rebound effect that have to be compensated by other sources later on, if not operated below the maximum power point (MPP)~\cite{wilches-bernalFundamentalStudyApplying2016}. 

In the Nordic grid, FCR is almost exclusively provided by hydropower. The controllability and storage capability of hydropower makes it ideal for this purpose. In recent years, however, the inertia reduction due to the renewable energy transition has made the bandwidth limitations associated with non-minimum phase (NMP) waterway dynamics a problem. Since the bandwidth of hydro-FCR cannot be increased without reducing the closed-loop stability margins \cite{agneholmFCRDDesignRequirements2019}, the Nordic SO's have developed a new market for FFR \cite{entso-eFastFrequencyReserve2019}. 
Units participating in FFR are subjected to ramp down limits and a \SI{10}{\second} buffer period before the device is allowed to recover energy exerted during the FFR event. This helps to avoid a secondary frequency dip before the hydro-FCR have fully activated. However, the requirement of a recovery-period disqualifies the use of uncurtailed {\WT}s. Since these operate at the MPP, any temporary power outtake will decelerate the turbine, thereby immediately lowering the sustainable power output. The open-loop control method proposed in \cite{entso-eFastFrequencyReserve2019} is therefore a potentially costly solution that require controllable storage devices such as batteries or curtailing of {\WT}s in order to guarantee the needed FFR capacity.

The contribution of this work is the design of a {closed-loop} decentralized dynamic ancillary service, distributing FCR (and FFR) between a heterogeneous ensemble of devices, to form a DVPP. The controllers rely on dynamic participation factors (DPFs) and are designed so that all devices collectively match the Bode diagram of a design target, specified by the SO requirements. 
Typically the design target will take the form of a low-pass filter that matches the desired steady-state FCR, but also models the allowed roll-off at higher frequencies. The frequency-dependent DPFs allows us to conveniently allocate the resources in a smart grid with many controllable actuators, and to account for the capacity and speed limitation of each device.
To validate our solution, we design a DVPP made up of hydro units, battery storage, and {\WT}s. We also show how the design can  improve the frequency stability of low-inertia power systems using a model of the Nordic synchronous grid. 
By ensuring that the DVPPs that participate in the FCR match a desired dynamic design target, the FCR requirements are achieved with slowest possible response speed and minimal input use. In comparison to related works \cite{liPowerElectronicInterface2008,glavinStandalonePhotovoltaicSupercapacitor2008,mullerLargescaleDemonstrationPrecise2019,morrenWindTurbinesEmulating2006,zhaoFastFrequencySupport2020,wilches-bernalFundamentalStudyApplying2016,entso-eFastFrequencyReserve2019}, this allows us to minimize the control effort, and thus avoid over-dimensioning of the devices that participate in the FCR by providing appropriate compensation for undesirable dynamic properties such as the NMP response of hydro units or the rebound effect of {\WT}s.

The remainder of the paper is structured as follows. \cref{sec:Prob_Form} presents the control problem, introducing the test system and models of controllable energy sources. \cref{sec:DVPP_control_design}, formally introduces the DVPP control design. In \cref{sec:ex_DVPP:ctrl:design} a local DVPP is designed and in \cref{sec:coordinated_in_N5} we close the frequency loop and coordinate FCR and FFR in the whole grid.   \cref{sec:Conclusion_Ensemble} concludes the paper with a discussion of the results.

%%%%%%%%%%%%%%%%%%%%%%%%%%%%%%%%%%%%%%%%

\section{Problem Formulation}
\label{sec:Prob_Form}
In this work, we are interested in the frequency containment and post fault dynamics of the center of inertia (COI).  That is, we do not directly address short term synchronization and inter-area oscillations in the control design. In the end, stability is verified in simulations by applying the control to a detailed power system model designed for large signal analysis. 

Power balance between production and consumption is ensured by controlling the COI frequency \cite{kundurPowerSystemStability1994}. For a network with $n_\mathrm{gen}$ synchronous machines, the COI frequency is 
\begin{equation}
    \omega_\coi = \frac{\sum_{i=1}^{n_\mathrm{gen}} M_i \omega_i}{M}, \quad M = \sum\nolimits_{i=1}^{n_\mathrm{gen}} M_i
\end{equation} 
where $\omega_i$ is the speed  and $M_i$ the  inertia of machine $i$.
Assuming that the grid stays connected, the motion of the COI frequency is determined by the  power balance
\begin{equation}
    sM \omega_\coi = P_\coi = \sum\nolimits_{i=1}^n P_{\mathrm{in},i} - P_{\mathrm{out},i}
\end{equation}
of the $n$ inputs and outputs distributed all over the system. 

For the analysis, we assume that physical frequency-dependent or frequency-controlled power sources can be linearized, e.g., neglecting effects of saturation.
The power balance $P_\coi$ is divided into frequency-dependent power sources  $F_i(s) \omega_\coi$ and external power sources $u_i$, so that
\begin{equation}
    P_\coi = \sum\nolimits_{i=1}^n F_i(s) \omega_\coi + u_i = F(s) \omega_\coi + u.
\end{equation}
We can then express the COI frequency disturbance response
\begin{equation}
\label{eq:coi_dist_simple}
    \omega_\coi = \frac{1}{s M + F(s) } u.
\end{equation}

Let $F_i(s)$ be broken up into $F_i(s) = D_i(s) + H_i(s) \cdot K_i(s)$, where $D_i(s)$ is some fixed frequency-dependent load or power source (typically assumed to be a constant), $H_i(s)$ represent the dynamics of some controllable power source, and $K_i(s)$ is a linear FCR controller taking a measurement of the local frequency as input.
The goal is then to design $K_i(s)$, $i\in\{1,\ldots,n\}$, so that \eqref{eq:coi_dist_simple} fulfills the FCR requirements of the SO. In this paper, we will study this problem using a case study of a 5-machine representation of the Nordic synchronous grid.

\subsection{The Nordic 5-Machine  Test System}
\label{sec:Nodic5_intro}
\begin{figure}[t!]
    \captionsetup[subfloat]{farskip=0pt}
    \centering
    \begin{minipage}[b][][b]{0.46\linewidth} 
    \subfloat[\label{fig:Nordic_5} One-line diagram.]
    {{\includegraphics[width=\textwidth]{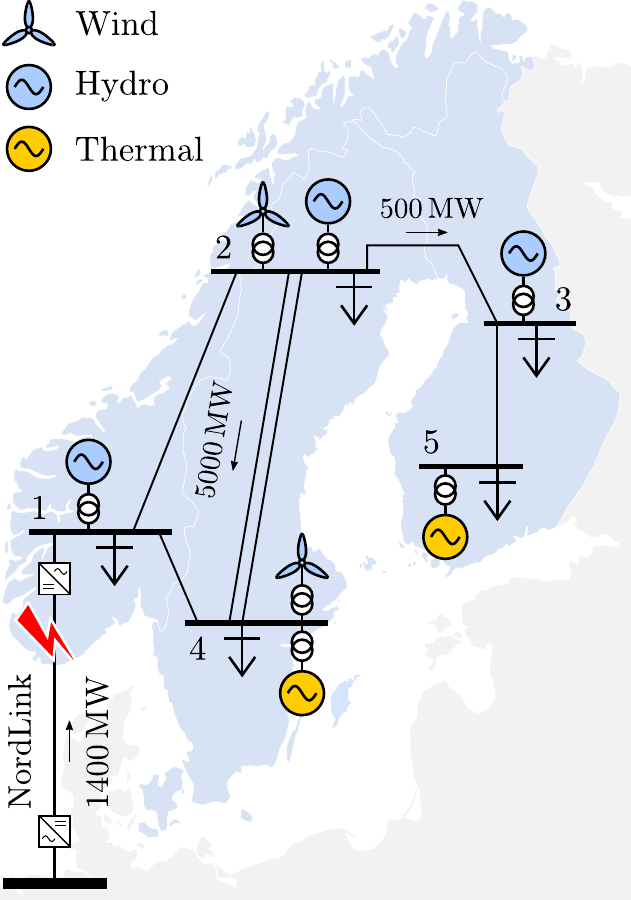}}}
    \end{minipage}
    \hfill
    \begin{minipage}[b][][b]{0.52\linewidth}
    \subfloat[\label{fig:Nordic5_step} Ideal FCR response.]
    {\makebox[\dimexpr\width+0.9cm\relax][c]{\includegraphics[scale=0.52]{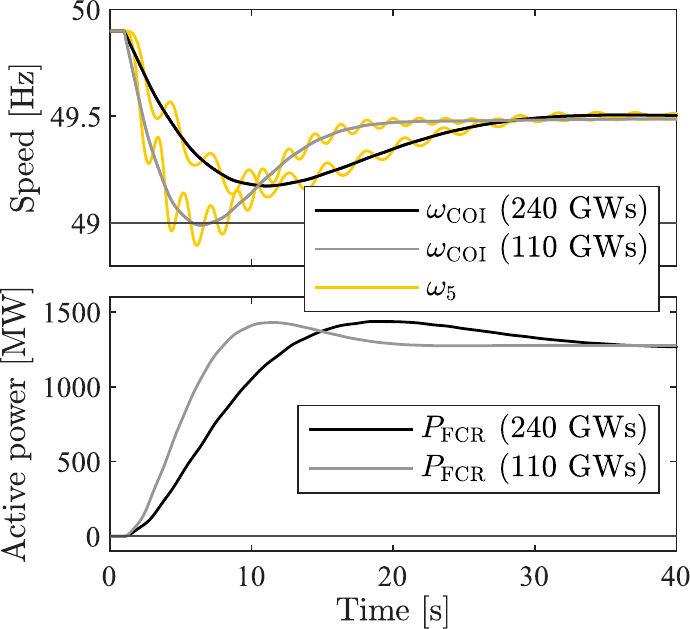}}}
    \\
    \vspace{10pt}
    \\
    \subfloat[\label{fig:Nordic5_bode} Bode diagram of the FCR open-loop.]
    {\makebox[\dimexpr\width+0.9cm\relax][c]{\includegraphics[scale=0.52]{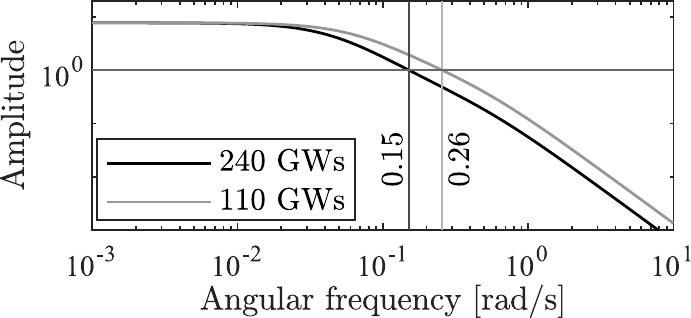}}}
    \end{minipage}
    \caption{The N5 test system. The full model, and test cases presented in this work, are available at the repository \url{https://github.com/joakimbjork/Nordic5}.
    }
    \label{fig:Nordic5_hydro_intro}
\end{figure}

Consider the Nordic 5-machine (N5) test system shown in \cref{fig:Nordic_5}. The system is phenomenological but has dynamical properties similar to the Nordic synchronous grid. The model is adapted from the empirically validated 3-machine model presented in \cite{saarinenFullscaleTestModelling2016}. 
Loads, synchronous machines and {\WT}s are lumped up into a single large unit at each bus.
The model is developed in Simulink Simscape Electrical \cite{hydro-quebecSimscapeElectricalReference2020_manual}. Hydro and thermal units are modeled as 16\textsuperscript{th} order salient-pole and round rotor machines, respectively. Assuming that inverters are operated within allowed limits and are fast enough so that their dynamics have only a marginal effect on \eqref{eq:coi_dist_simple}, we model all inverter sources as grid-following controllable power loads.

The amount of synchronous generation connected to the grid varies with the load demand and dispatch. Therefore, the amount of system kinetic energy varies greatly over the year \cite{entso-eFastFrequencyReserve2019}. 
Here, we will consider a high-inertia scenario, adapted from \cite{saarinenFullscaleTestModelling2016}, with $W_\mathrm{kin} = \SI{240}{\giga\wattsecond}$ and a low-inertia scenario with $W_\mathrm{kin} = \SI{110}{\giga\wattsecond}$ distributed according to \cref{tab:Nordic5}. Loads are modeled as constant power loads with a combined proportional frequency dependency $D(s) = D =  \SI{400}{\mega\watt\per\hertz}$. 

\begin{table}[t!]
\centering
\caption{Machine parameters for the \SI{240}{\giga\wattsecond} and \SI{110}{\giga\wattsecond} test cases.}
\label{tab:Nordic5}
\begin{tabular}{c|cc|cc}
\hline
  Bus & 
  $W_\mathrm{kin}$ [\si{\giga\wattsecond}] & $P_e$ [\si{\mega\watt}] & $W_\mathrm{kin}$ [\si{\giga\wattsecond}] & $P_e$ [\si{\mega\watt}]  
  \\
  \hline
 1 & 67.5 & \SI{18000}{} & 34 & \SI{9000}{}
\\
2 & 45 & \SI{12000}{} & 22.5 & \SI{6000}{}
\\
3 & 7.5 & \SI{2000}{}  & 7.5 & \SI{2000}{}  
\\
4 & 73.3 & \SI{11000}{} & 33 & \SI{5000}{} 
\\
5 & 46.7 & \SI{7000}{}  & 13 & \SI{2000}{}  
\\
\hline
$\Sigma$&  240 & & 110 &
\\
\hline
\end{tabular}
\end{table}

To specify a desired ``ideal" FCR response, we use the FCR for disturbance (FCR-D) specifications in the Nordic synchronous grid. 
The FCR-D is used to contain the frequency outside normal operation. 
Following a rapid frequency fall from 49.9 to \SI{49.5}{\hertz}, the reserves should be \SI{50}{\percent} activated within \SI{5}{\second} and fully activated in \SI{30}{\second}.
Following larger disturbances the maximum instantaneous frequency deviation (the nadir) should be limited to \SI{1.0}{\hertz} \cite{entso-eNordicSynchronousArea2018}.
Hence, we let the FCR-D design target take the form
\begin{equation}
\label{eq:asmFCRD}
    F_\mathrm{FCR}(s) =  R_\mathrm{FCR} \frac{6.5s+1}{(2s+1)(17s+1)}.
\end{equation}
Consider the dimensioning fault to be the instant disconnection of the {NordLink dc cable} \cite{NordLink_manual} importing \SI{1400}{\mega\watt} from Germany into Norway as shown in \cref{fig:Nordic_5}.
Choosing $R_\mathrm{FCR} = \SI{3100}{\mega\watt\per\hertz}$, the post-fault system stabilizes at \SI{49.5}{\hertz}, as seen in \cref{fig:Nordic5_step}. The second-order filter in \cref{eq:asmFCRD} is tuned so that the FCR-D requirements are fulfilled for both scenarios,  while also avoiding an overshoot and a second frequency dip when the frequency is restored. 

\begin{figure}[t]
	\centering
	\includegraphics[width=0.9\linewidth]{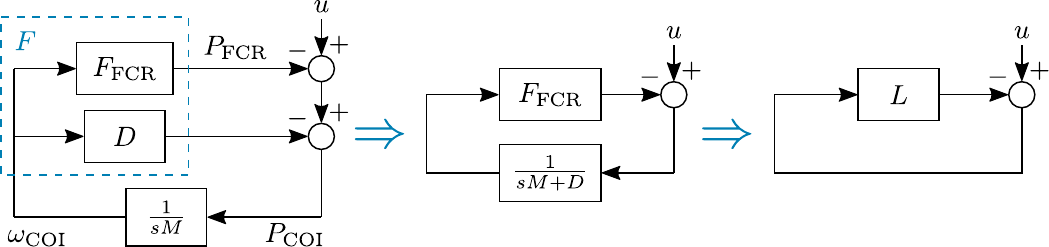}
	\caption{Block diagram of the FCR control loop.}
	\label{fig:block_coi}
\end{figure}

In \cref{fig:Nordic5_step,fig:Nordic5_bode}, we consider ideal actuation $H_i(s) =1$. Thus, \eqref{eq:asmFCRD} is realized with ideal controllable power sources distributed at buses 1, 2, and 3 so that the total controlled input $P_\mathrm{FCR} = F_\mathrm{FCR}\s(\omega_\mathrm{ref}-\hat\omega)$, where $\omega_\mathrm{ref}$ is the frequency reference and $\hat\omega\approx \omega_\mathrm{COI}$ is the locally measured frequency. As shown in \cref{fig:Nordic5_step}, this approximation has no big impact on the result, assuming that the post fault system remains stable. 
With reduced inertia, the speed of the system increases. This also increases the cross-over frequency of the FCR open-loop
\begin{equation}
\label{eq:FCR_open_loop}
    L(s) =  F_\mathrm{FCR}(s) \frac{1}{sM+D},
\end{equation}
obtained by breaking the loop at the input/output of $F_\mathrm{FCR}(s)$, as shown in \cref{fig:block_coi}.
Since real actuators will have bandwidth limitations, the low-inertia scenario therefore poses a greater control challenge. 

When deviating from the above ideal actuation scenario,
as we will see, the NMP characteristics of hydro units will make it impossible to match the design target \cref{eq:asmFCRD}.
The target can be modified, by increasing the cross-over frequency, so that the FCR-D requirements are fulfilled even if FCR are delivered by hydro governors. However, due to bandwidth limitations imposed by the NMP zeros, see next section for details, this is not a good solution since this reduces the closed-loop stability margins  \cite{agneholmFCRDDesignRequirements2019}. Because of this, the Nordic SOs have developed a new market for FFR \cite{entso-eFastFrequencyReserve2019}. FFR can be provided by, e.g., battery sources or wind farms bidding on such a market. 

In this work, we consider the control problem of coordinating multiple heterogeneous plants with different time constants and limitations. We will consider hydro units, batteries, and {\WT}s.

\subsection{Hydro Governor Model}
\label{sec:hydro_model}
\begin{figure}[t!]
		\centering
		\includegraphics[width=1\linewidth]{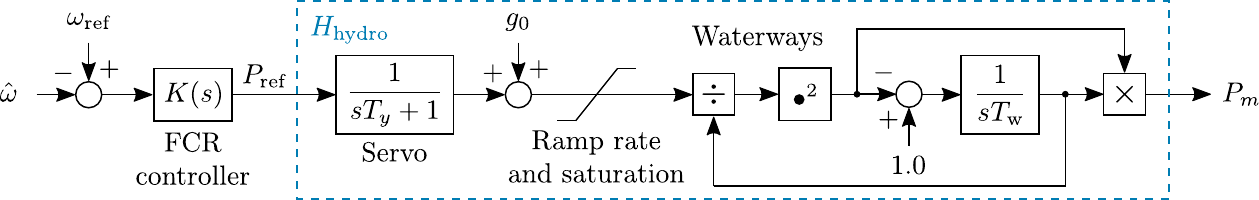}
		\caption{Block diagram of the hydro turbine and governor model.}
		\label{fig:hydro_block}
\end{figure}

The hydro governor model implemented in this work is an adaption of the hydro governor model available in the Simulink Simscape Electrical library \cite{hydro-quebecSimscapeElectricalReference2020_manual};
modified to allow a general linear FCR controller, $K(s)$, instead of the predefined PID/droop control structure as shown in \cref{fig:hydro_block}. The nonlinear second-order model is useful for large-signal time-domain simulations.
For the linear design, the turbine is modeled as
\begin{equation}
\label{eq:Hhydro}
    H_\textrm{hydro}(s) = 
     2\frac{ z - s}{s + 2z}  
    \frac{1}{sT_y + 1} 
    , \quad z = \frac{1}{g_0 T_\mathrm{w}},
\end{equation}
where $T_y$ is the servo time constant, $g_0$ the initial gate opening, and $T_\mathrm{w}$ the water time constant \cite{kundurPowerSystemStability1994}. 

Following a gate opening, the pressure over the turbine falls before the water accelerates, due to the inertia in the water column. Because of this, the initial power surge will be in the opposite direction of the gate opening change. This behaviour results in a bandwidth limitation which in the linearized model \cref{eq:Hhydro} is characterized by the RHP zero \cite{kundurPowerSystemStability1994}. 

\subsection{Battery Storage Model}
In the time frame of interest for frequency control, the dynamics of battery storage units are dominated by the dynamics of the inverter and its controls \cite{alhelouPrimaryFrequencyResponse2020,schifferSynchronizationDroopcontrolledMicrogrids2013}. Assuming that the inverter dynamics have no significant impact on \cref{eq:coi_dist_simple}, see \cref{rem:rule_of_thumb} later on, we therefore model batteries as ideal controllable power sources, with 
\begin{equation}
    H_\mathrm{battery}(s) = 1.
\end{equation}
For the simulation case study we also keep track of the energy level to indicate the required battery size. Depending on the size of the energy storage, batteries can be used as both FCR and FFR. In this work, we consider that the storage is limited so that batteries are used only for FFR.

\subsection{Wind Power Model}
\label{sec:wind}
\begin{figure}[t!]
		\centering
		\includegraphics[width=\linewidth]{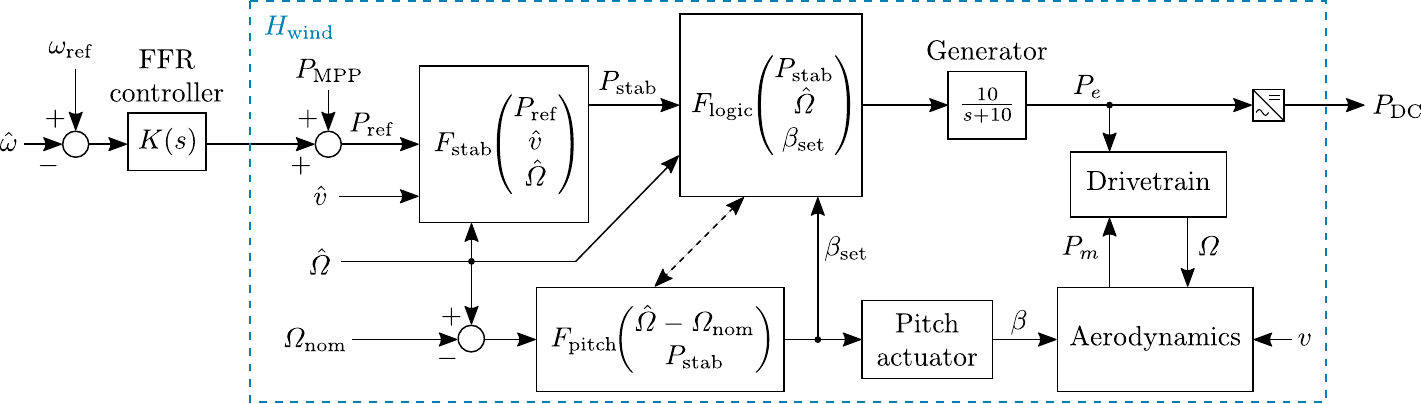}
		\caption{Block diagram of a variable speed controlled {\WT} \cite{bjorkVariablespeedWindTurbineunpublished}.}
		\label{fig:wind_block}
\end{figure}

We consider a 8\textsuperscript{th} order {\WT} model shown in \cref{fig:wind_block}. The model is based on the 
National Renewable Energy Laboratory (NREL)  \SI{5}{\mega\watt} baseline {\WT} model \cite{jonkmanDefinition5MWReference2009}. The control system has been modified by adding a stabilizing feedback controller, $F_\mathrm{stab}$, to allow the turbine to participate in FFR. For a full description and analysis of the modified turbine, the reader is referred to \cite{bjorkVariablespeedWindTurbineunpublished}. Here we give a brief overview of the {\WT} model and its linearization.

The pitch controller, $F_\mathrm{pitch}$, ensures that the rotor speed $\varOmega$ does not exceed the rated speed $\varOmega_\mathrm{nom}$, by adjusting the pitch angle $\beta$. The control logic, $F_\mathrm{logic}$ allows the turbine to operate in various operating modes by adjusting the generator set-point and choosing when to activate the pitch controller. 

Assume uncurtailed operation at the MPP below the rated wind speed, then $\beta = 0$ and $P_e = P_\mathrm{MPP}$. Assuming that the inverter dynamics have no significant impact on \cref{eq:coi_dist_simple}, we let $P_\mathrm{in}=P_\mathrm{DC} = P_e$.
The mechanical power $P_m$ is a function of rotor speed $\varOmega$ and the wind speed $v$. Any deviation from the optimal speed $\varOmega_\mathrm{MPP}$ will result in a reduced sustainable power output.
However, if operated below rated speed, the electric power can be temporarily increased, allowing the {\WT} to participate in FFR. This however will decelerate the rotor and reduce the sustainable power output, as shown by the power/speed characteristics in \cref{fig:wind_power-speed} later on.
To ensure stability, a variable speed feedback controller, $F_\mathrm{stab}$, is implemented \cite{bjorkVariablespeedWindTurbineunpublished}. The controller uses measurements $\hat v$ and $\hat \varOmega$ of the wind and rotor speed, to modify the power reference $P_\mathrm{ref}$ to $P_\mathrm{stab}$. 
As shown in \cite{bjorkVariablespeedWindTurbineunpublished}, the dynamics most relevant for FFR are
\begin{equation}
    P_e \approx \frac{s-z}{s + k_\mathrm{stab} - z} P_\mathrm{ref},
\end{equation}
where the RHP zero $z$ is a function of the drivetrain and aerodynamics, and $k_\mathrm{stab}$ is the effective stabilizing feedback gain from $F_\mathrm{stab}$ at the current wind speed. As the turbine decelerates, $z$ increases. Keeping the turbine above the minimum allowed speed, then $z \leq \bar z$.  Let $k_\mathrm{stab} = 2\bar z$, then
\begin{equation}
\label{eq:Hwind}
    H_\mathrm{wind}(s) = \frac{s-\bar z}{s + \bar z}
\end{equation}
is a linear representation useful for analysis and control design. For the modified NREL turbine,  $\bar{z} =  5.8 v\cdot 10^{-3}$ \cite{bjorkVariablespeedWindTurbineunpublished}.

%%%%%%%%%%%%%%%%%%%%%%%%%%%%%%%%%%%%%%%%%%%%%%%%%%%%%%%%%%%%

\section{DVPP Control Design}
\label{sec:DVPP_control_design}
Using FFR from wind to assist FCR from slower conventional generation has been proposed in the literature \cite{morrenWindTurbinesEmulating2006,wilches-bernalFundamentalStudyApplying2016,zhaoFastFrequencySupport2020}. In this work, we take this one step further. We develop a linear control design method that coordinates the dynamic response of a heterogeneous ensemble of plants, so that the combined Bode plot of all participating plants matches a target function over all frequency ranges. Using the target function  \eqref{eq:asmFCRD} result is a DVPP that meets the SO's FCR-D requirements. The method is general and allows us to take into account energy capacity, power, and bandwidth limitations. From a control design perspective, the limitations imposed by NMP zeros will be the hardest to address, since they also affect the stability margins. Therefore, this work focuses on combining hydro and wind. 

This section presents a coordinated FCR and FFR control design method. It can be applied globally, or locally in a DVPP. Our design is based on the COI model and assumes asymptotic synchronization on the average mode. 
In the end, stability is verified by applying the control and simulating the power system model.

\subsection{Coordinated FCR and FFR Using Model Matching}
Let $D\s = D$ represent the uncontrolled, proportional, frequency dependent loads in the system and let $\bm H\s = 
[
H_1\s,\ldots, H_n\s
]^\T$, and $\bm K\s = [
K_1\s,\ldots, K_n\s
]^\T$.

\begin{figure}[t]
	\centering
	\includegraphics[scale=0.9]{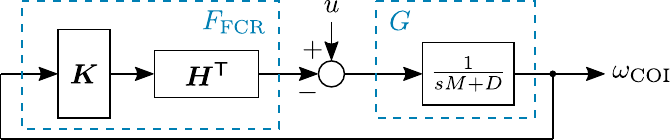}
	\caption{Block diagram of the FCR design problem.}
	\label{fig:block_internal_stab}
\end{figure}

Breaking the loop at the input of $\bm K\s$ in \cref{fig:block_internal_stab} the \emph{global} open-loop gain of the FCR control scheme
becomes
\begin{equation}
\label{eq:open_loop_def}
L\s = \sum\nolimits_{i=1}^{n} L_i\s, \quad L_i\s = G\s \cdot H_i\s \cdot  K_i\s
\end{equation}
where $G\s = {1}/({sM + D})$.

We pose the DVPP design as a decentralized model matching problem.
Let $\Lcoides\s = G\s \cdot F_\mathrm{FCR}\s$ be the desired, stable and MP, loop-gain that fulfill the FCR-D specifications. The loop-gain of each plant is then given by
\begin{equation}
\label{eq:L_i}
L_i\s = c_i\s \cdot \Lcoides\s = c_i\s \cdot G\s \cdot F_\mathrm{FCR}\s
\end{equation}
where $c_i\s$, are DPFs to be designed. DPFs are frequency-dependent version of static participation factors \cite{kundurPowerSystemStability1994}, which allow us to take the dynamic characteristics of each device into account.
The controller for subsystem $i$ is then given by
\begin{equation}
K_i\s=  c_i\s \cdot {F_\mathrm{FCR}\s }/{H_i\s}.
\end{equation}	

We say that perfect model matching is achieved if 
\begin{equation}
\label{eq:model_matching}
\sum\nolimits_{i=1}^n c_i(s) = 1, \quad \forall s\in \complex
\end{equation}
in which case $L_1\s+\cdots+L_n\s = c_1\s \cdot \Lcoides\s +\cdots+c_n\s \cdot \Lcoides \s = \Lcoides\s$.
However, taking into account the dynamical constraints of the actuators $H_i\s$, such as NMP zeros, we may have to relax \eqref{eq:model_matching} to ensure that $K_i\s$ is proper and that the resulting closed-loop system is internally stable. We relax \eqref{eq:model_matching} by allowing a mismatch in the equality.
Typically, we want a good match at steady state up to some frequency $\omega_B$, e.g., we want $\sum_{i=1}^n c_i(\jomega) = 1$ for $\omega \in [0,\omega_B]$. 

\begin{rem}
\label{rem:rule_of_thumb}
A rule of thumb is that, for the resulting closed-loop system to be close to nominal, we want a good match up until at least ten times the cross-over frequency \cite{skogestadMultivariableFeedbackControl2007}. For model matching in the low-inertia N5 test case in \cref{fig:Nordic5_bode} this indicates that $\omega_B \approx \SI{2.6}{\radian\per\second}$. This also justifies neglecting stable dynamics of batteries and power electronics in the FCR and FFR control design since these typically are $\gg \SI{2.6}{\radian\per\second}$ \cite{alhelouPrimaryFrequencyResponse2020,schifferSynchronizationDroopcontrolledMicrogrids2013}.
\end{rem}

\subsection{Internal Stability}
In addition to shaping the COI frequency disturbance response, we have to ensure internal stability with respect to the interfaces between the plant $G\s$, the FCR controllers $K_i\s$, and the controllable power sources $H_i\s$ in \cref{fig:block_internal_stab}.

\begin{thm}[\hspace{1sp}\cite{zhouRobustOptimalControl1996}]
\label{thm:intrenal_stab}%
	The system is internally stable if and only if the sensitivity 
	\begin{equation}
	\label{eq:sens}
	S\s = \frac{1}{1 + L\s}
	\end{equation}
	 is stable and no unstable pole-zero cancellations occur between plants and controllers $G\s$, $H_i\s$, and $K_i\s$, $i\in \{1,\ldots,n\}$.
\end{thm}

\begin{cor}%
	In case of RHP  poles  $p_j \in \complex_+$ in $H_i\s$ or $G\s$ we need that 
	\begin{equation}
	L_i^\inv(p_j) =  0.
	\end{equation}
	Since we are not allowed to cancel RHP poles, any RHP poles must therefore remain in the global open-loop $L\s$. However, if  $p_j$ belongs to $H_i\s$, we may pre-stabilize $H_i\s$ by first designing a local feedback controller \cite{patesRobustScaleFreeSynthesis2019}.
\end{cor} 

\begin{cor}
\label{cor_NMP_required}%
	In the case of NMP zeros $z_j\in \complex_+$ in $H_i\s$ or $G\s$ we need that 
	\begin{equation}
	L_i(z_j) = 0.
	\end{equation}
	Zeros cannot be moved by series compensation or feedback. So unlike unstable poles, these must remain in the system.
	However, since zeros are moved by parallel connections, as in \cref{fig:block_internal_stab,fig:DVPP_block}, it is not necessary for the NMP zeros of $H_i\s$ to remain in the global loop-gain $L\s$.
\end{cor}

For the remainder, we assume that $G\s$ is stable and MP, and that any unstable poles in $H_i\s$ have been pre-stabilized.
The problem that remains is then how to deal with NMP zeros. Ideally, we want the global open-loop to be MP so that perfect matching \eqref{eq:model_matching} can be achieved.

\subsection{Choosing Dynamic Participation Factors (DPFs)}
\label{sec:part_factor_algorithm}
There are many ways of choosing the DPFs $c_i(s)$, $i \in \{1,\ldots,n\}$.
Ideally, the factors are distributed between VPP units to play on their dynamic strengths, compensate for their weaknesses, and align with economic considerations. To illustrate how this can be achieved, we here propose a method where the frequency control is divided up into slower FCR and faster FFR.

Let $c_i(s)$, $i \in \{1,\ldots,m\}$ and $c_i(s)$, $i \in \{m+1,\ldots,n\}$ be the DPFs for  FCR and FFR, respectively. Let each producer specify a variable $k_i$ indicating their willingness or marginal cost for supplying FCR and FFR. Normalize the constants so that $\sum_{i=1}^m k_i = 1$ and $\sum_{i=m+1}^n k_i = 1$.

Starting with FCR, let
\begin{equation}
\label{eq:FCR_part_fact}
    c_i(s) = k_i {\mathcal{B}_i(s)}/{\mathcal{B}_i(0)}, \quad i\in\{1,\ldots,m \}
\end{equation}
where, as necessary according to \cref{cor_NMP_required},
\begin{equation}
\label{eq:blaschke}
    \mathcal{B}_i(s) = \prod\nolimits_{j=1}^{n_z} \frac{z_j-s}{s+p_j}
\end{equation}
contains all $n_z$ NMP zeros of the plant $H_i(s)$. 
The poles $p_j$ are design parameters, e.g., to adjust the cross-over frequency of $L_i(s)$. A good starting point however, is to let $p_j = z_j$ so that the DPFs are all-pass.

Next, we design the FFR participation factors. 
Let
\begin{equation}
    c_i(s) = k_i \frac{\mathcal{B}_i(s)}{\mathcal{B}_i(\infty)} \left(1-\sum\nolimits_{l=1}^m c_l(s)\right) , \quad i\in\{m+1,\ldots,n \}
\end{equation}
{where $\mathcal{B}_i(s)$ is the product \cref{eq:blaschke}. Note that  $\mathcal{B}_i(\infty)$ is a negative real number if $n_z$ is odd.}

At this point, we have FCR and FFR controllers that achieves perfect matching $\sum_{i=1}^n c_i(s)=1$ for $s=0$ and $s\rightarrow \infty$. Since no NMP zeros are cancelled, internal stability is achieved if the sensitivity \eqref{eq:sens} is stable. However, if any of the FFR plants $H_i(s)$ are NMP, we do not have perfect model matching, due to the required modifier ${\mathcal{B}_i(s)}/{\mathcal{B}_i(\infty)}$. If the total sum  is MP however, this can be amended by adding a final normalization step $c_i'(s) = c_i(s)/\sum_{i=1}^n c_i(s)$.

\section{Illustrative Examples on DVPP Control Design}
\label{sec:ex_DVPP:ctrl:design}
In this section, we will show how a set of heterogeneous plants can be controlled so that they together form a DVPP with favourable MP characteristics. We do this, using DPFs as described in \cref{sec:DVPP_control_design}. For simplicity, we consider open-loop control of a subsystem connected to the grid. Therefore, it is not yet possible to state any requirement on the loop gains \cref{eq:open_loop_def,eq:L_i}. Instead, the design is specified in terms of the ideal FCR response \eqref{eq:asmFCRD}. The insight gained from this will later be used for the Nordic case study.

\subsection{FCR Provided by two Hydro Units}
\label{sec:hydropower_FCR}
\begin{figure}[t!]
		\centering
		\includegraphics[scale=0.9]{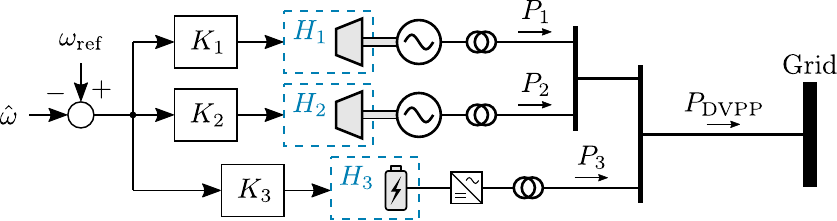}
		\caption{
		 One-line diagram of a battery-hydro DVPP.
		}
		\label{fig:DVPP_block}
\end{figure}

% 	See \cref{proof:sum_is_NMP}. \phantom\IEEEQEDhere
Consider a subsystem with two \SI{50}{\mega\voltampere} hydro units exporting power to the grid as shown in \cref{fig:DVPP_block}. Assume water time constants $T_\mathrm{w,1} = \SI{1.25}{\second}$ and $T_\mathrm{w,2} = \SI{2.5}{\second}$, respectively, and an initial gate opening $g_0 = 0.8$ and servo time constant $T_y = \SI{0.2}{\second}$ for both turbines, then
\begin{equation}
    H_1\s =  2 \frac{-s + 1}{s+2} \frac{1}{s 0.2 + 1}, \ H_2\s =  2 \frac{-s + 0.5}{s+1}\frac{1}{s 0.2 + 1}.
\end{equation}

\subsubsection*{Goal} Design FCR controllers $K_1\s$ and $K_2\s$ for the two-hydro subsystem so that: both units increase their steady-power output by \SI{10}{\mega\watt} following \SI{1}{\hertz} frequency reference step, and so that $H_1\s \cdot K_1\s + H_2\s \cdot K_2\s  $ comes close to the ideal FCR response \eqref{eq:asmFCRD}
with $R_\mathrm{FCR} = \SI{20}{\mega\watt\per\hertz}$.

\subsubsection*{Solution} 
From \cref{prop:sum_is_NMP} in the Appendix, we know that the two-hydro subsystem will have a RHP zero $z \in [0.5,1]$. Thus, perfect matching is not realizable. 
The design criteria requires that $c_1(0) = c_2(0) = 0.5$, and for internal stability,  RHP zeros need to be included. Let $K_i\s = c_i\s \cdot F_\mathrm{FCR}\s/H_i\s$, $i \in\{1,2 \}$ with
\begin{equation}
\label{eq:hydro_part_factors}
        c_1\s =0.5 \frac{-s + 1}{s+1} \quad \text{and} \quad c_2\s =0.5 \frac{-s + 0.5}{s+0.5}.
    \end{equation}
    
\subsubsection*{Result}
The resulting sum equals
\begin{equation}
    c_1\s+c_2\s = \frac{(-s+1/\sqrt{2})(s+1/\sqrt{2})}{(s+1)(s+0.5)}.
\end{equation}
\begin{figure}[tb!]
    \captionsetup[subfloat]{farskip=0pt}
    \centering
    \subfloat[\label{fig:bode_two_hydro}Participation factors.]
    {{\includegraphics[scale=0.52]{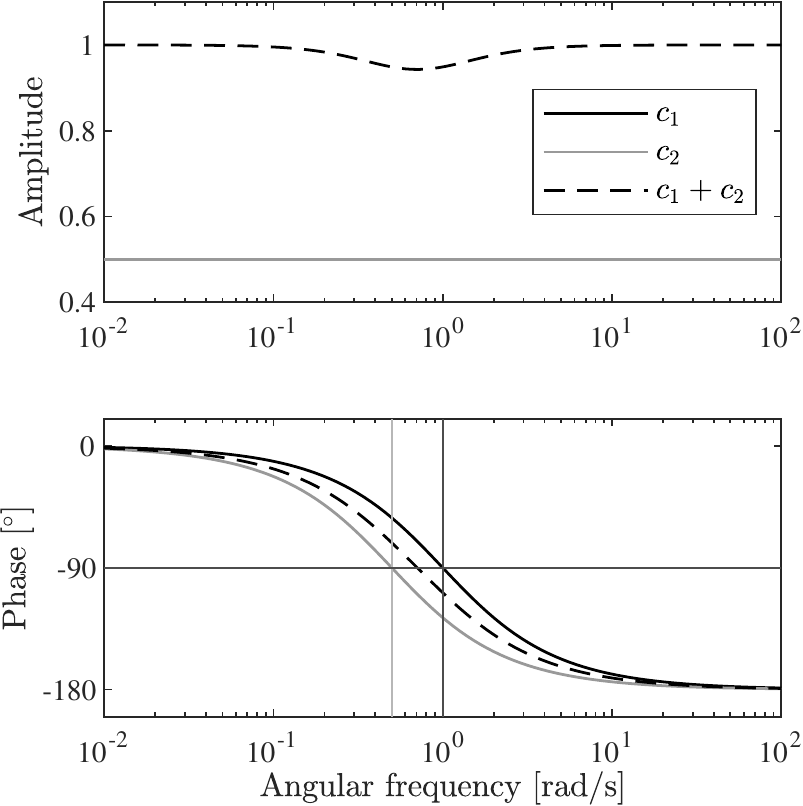}}}
    \hfill
    \subfloat[\label{fig:DVPP_hydro_lines}Power injections.]
    {{\includegraphics[scale=0.52]{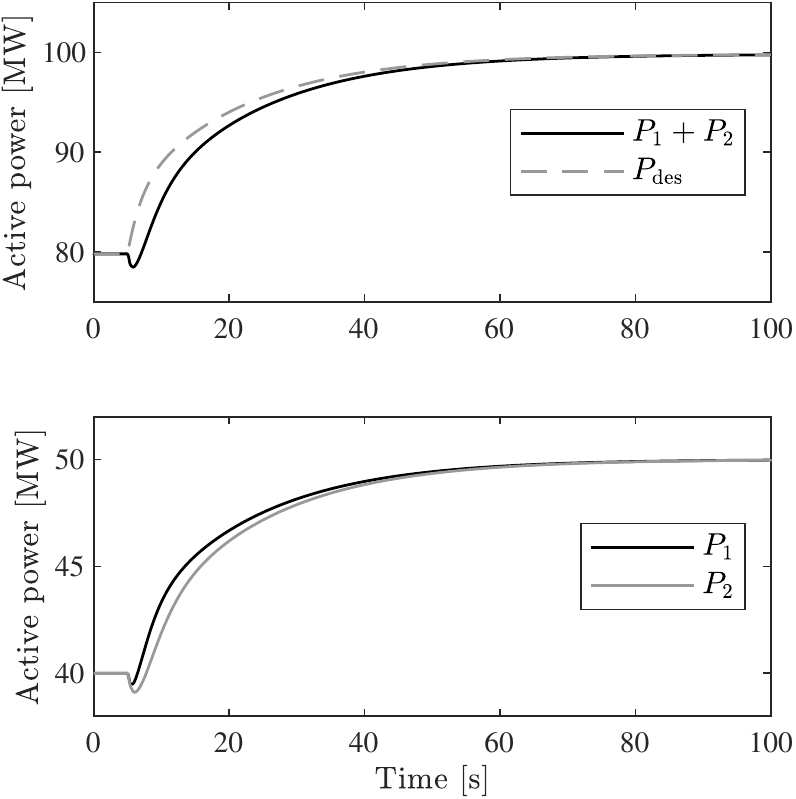}}}
    \caption{Power output and bode diagram of the two hydro units.}
    \label{fig:DVPP_hydro}
\end{figure}

As shown in \cref{fig:bode_two_hydro} the model matching is fulfilled at steady-state with $c_1(0)+c_2(0) = 1$. However, the consequence of the RHP zero is that the DPF $c_1\s+c_2\s$ have a \SI{-180}{\degree} phase shift at higher frequencies.  
The \SI{1}{\hertz} step response in \cref{fig:DVPP_hydro_lines} shows the characteristic NMP initial drop.
This is an unavoidable physical property of the hydro governors that makes it impossible to fulfill the design target \eqref{eq:asmFCRD}.

Although perfect matching is impossible, the performance can be improved by modifying the participation factors, e.g., by selecting faster poles in \cref{eq:hydro_part_factors}, or by selecting a design target \eqref{eq:asmFCRD} with a higher cross-over frequency. However, in low-inertia power systems, this may result in dangerously low closed-loop stability margins or even instability \cite{agneholmFCRDDesignRequirements2019}. A more robust solution is to complement hydropower with FFR.

%%%%%%%%%%%%%%%%%%%%%%%%%%%%%%%%%%%%%%%

\subsection{Battery Storage FFR Support}
As shown in \ref{sec:hydropower_FCR}, it is impossible to achieve a MP FCR response using only hydro units. To improve the transient response, one solution is to assist the hydro units with FFR from a battery storage, forming a DVPP as shown in \cref{fig:DVPP_block}.

\subsubsection*{Goal}
Consider the battery storage as an ideal controllable power source, with $H_3(s) = 1$. Design $K_3(s)$ so that the DVPP, with $K_i\s = c_i\s \cdot F_\mathrm{FCR}\s/H_i\s$, fulfills
\begin{equation}
\label{eq:fulfill_Ffcr}
    \sum\nolimits_{i=1}^3 K_i\s \cdot H_i\s = F_\mathrm{FCR}\s,
\end{equation}
that is, design the DPFs as in \eqref{eq:model_matching}.

\subsubsection*{Solution}
From \cref{prop:NMP_assisted_by_MP} in the Appendix, we know that since $\Real \big[c_1(\jomega) + c_2(\jomega)\big] \leq 0$, $\forall \omega$, the complementary DPF
\begin{equation}
\label{eq:battery_part_factor}
    c_3\s = 1-\big(c_1\s+c_2\s \big) = 2s\frac{(s+0.75)}{(s+1)(s+0.5)}
\end{equation}
is guaranteed to be stable and MP.
With the battery dynamics, $H_3\s$, being stable and MP, perfect matching is achieved with $K_3\s = c_3\s \cdot F_\mathrm{FCR}\s/H_3\s$.

\subsubsection*{Result}
\begin{figure}[tb!]
    \captionsetup[subfloat]{farskip=0pt}
    \centering
    \subfloat[\label{fig:bode_hydro_battery}Battery-hydro DVPP.]
    {{\includegraphics[scale=0.52]{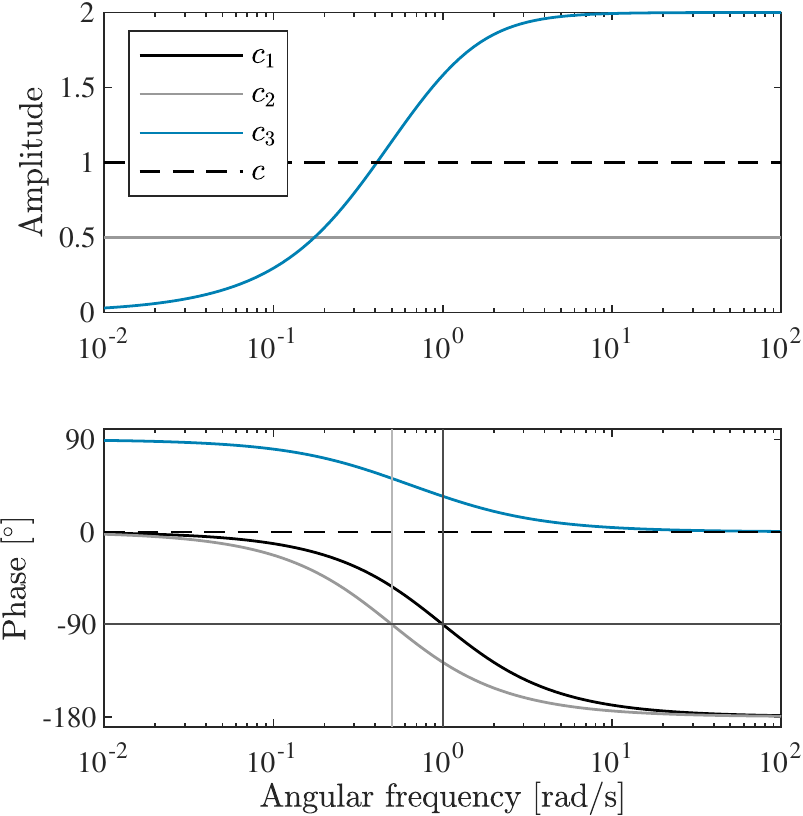}}}
    \hfill
    \subfloat[\label{fig:bode_hydro_wind}Wind-hydro DVPP.]
    {{\includegraphics[scale=0.52]{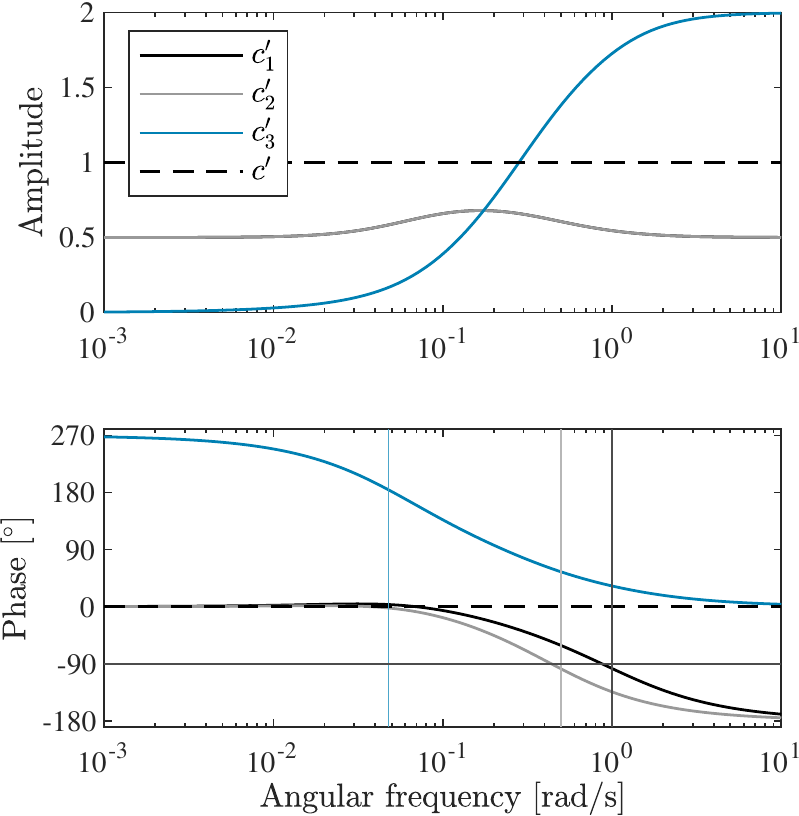}}}
    \caption{Bode diagram of participation factors, where  $c\s=\sum_{i=1}^nc_i\s$.}
    \label{fig:bode_DVPP}
\end{figure}

\begin{figure}[tb!]
    \captionsetup[subfloat]{farskip=0pt}
    \centering
    \subfloat[\label{fig:DVPP_storage_lines}Power injections.]
    {{\includegraphics[scale=0.52]{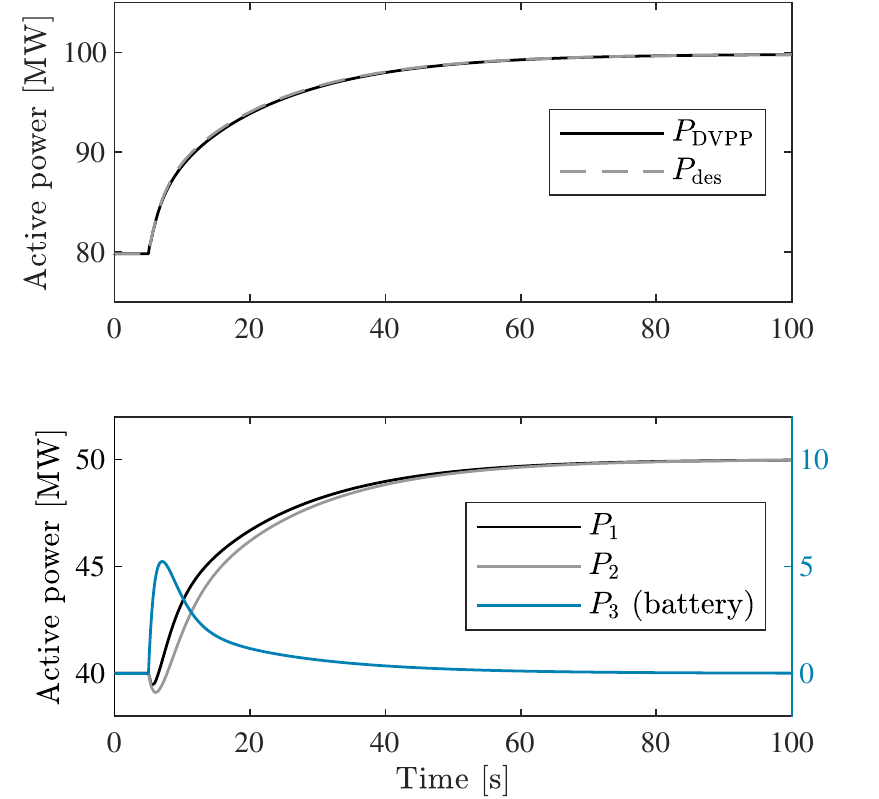}}}
    \hfill
    \subfloat[\label{fig:DVPP_storage}Battery storage.]
    {{\includegraphics[scale=0.52]{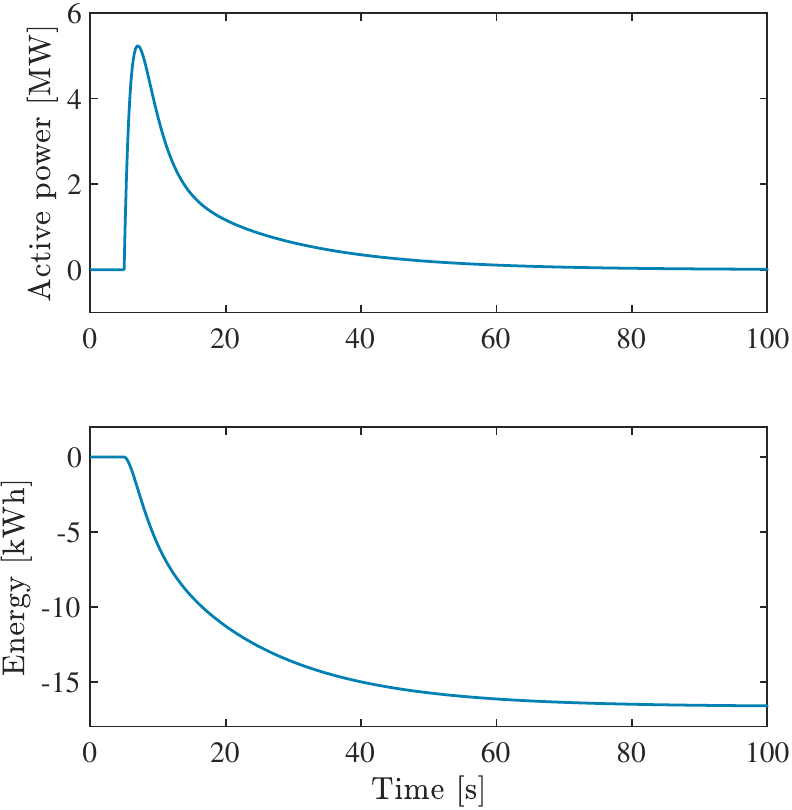}}}
    \caption{Power output of the battery-hydro DVPP.}
    \label{fig:DVPP_battery-hydro}
\end{figure}

As seen in \cref{fig:bode_hydro_battery}, the battery compensates for the phase lag of the hydro units so that $\sum_{i=1}^n c_i\s  = 1$. As a result, the DVPP output in \cref{fig:DVPP_storage_lines} matches the ideal response
$P_\mathrm{des} = F_\mathrm{FCR}\s (\omega_\mathrm{ref}-\hat\omega)$.
With a more detailed battery plus converter model, perfect matching can only be expected up to a certain frequency.

Having a stable MP controllable power source is ideal for providing FFR in a DVPP. With a battery storage, bounds on achievable performance are determined by the power rating and the storage capacity. As shown in \cref{fig:DVPP_storage}, for assisting the \SI{100}{\mega\voltampere} hydro park, we need at least \SI{5.5}{\mega\watt} and \SI{17}{\kilo\watthour}. 

%%%%%%%%%%%%%%%%%%%%%%%%%%%%%%%%%%%%%%%%%%%%%

\subsection{Wind Power FFR Support}
\label{sec:wind_power_supported}
An alternative to providing FFR with battery storage, is to assist with FFR from wind power. Compared to the battery solution however, for the {\WT}, its dynamics cannot be neglected. 
Here, we consider {\WT}s operated at the MPP, but below the rated wind speed so that the power output can, at least temporarily, be increased. 
When commanded by the FFR controller to exert power above the MPP, the rotor decelerates, and thus the sustainable power output decreases, as shown by the power/speed characteristics in \cref{fig:wind_power-speed}. The decline in sustainable power manifests in a NMP behavior \cite{bjorkVariablespeedWindTurbineunpublished}. 

\subsubsection*{Goal}
\begin{figure}[tb!]
    \captionsetup[subfloat]{farskip=0pt}
    \centering
    \subfloat[\label{fig:block_wind-hydro}One-line diagram of the DVPP.]
    {\makebox[\dimexpr\width+0.8cm\relax][c]{\includegraphics[scale=0.9]{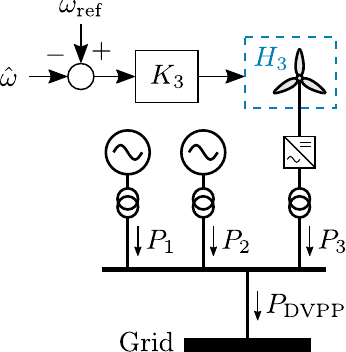}}}
    \hfill
    \subfloat[\label{fig:wind_power-speed}Wind farm power/speed characteristic.]
    {\makebox[\dimexpr\width+0.0cm\relax][c]{\includegraphics[scale=0.8]{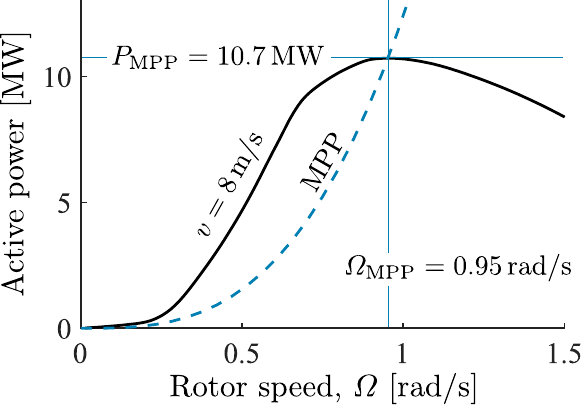}}}
    \caption{Wind-hydro DVPP with a \SI{30}{\mega\watt} wind farm.}
    \label{fig:DVPP_with_wind}
\end{figure}

Consider a \SI{30}{\mega\watt} wind park connected to the subsystem as shown in \cref{fig:DVPP_with_wind}. Assume initially that $P_3 = P_\mathrm{MPP}$.
Let the wind speed be $v=\SI{8}{\meter\per\second}$. Then the all-pass filter
% , $P_\mathrm{MPP} = \SI{10.7}{\mega\watt}$, and
\begin{equation}
\label{eq:H3_wind}
    H_3\s = \frac{s - 0.048}{s+0.048}
\end{equation}
is a suitable linear representation of the {\WT} dynamics \cite{bjorkVariablespeedWindTurbineunpublished}.
Design $K_3\s$ so that the DVPP fulfills \eqref{eq:fulfill_Ffcr}.

\subsubsection*{Solution} 
Using the three-step approach in \cref{sec:part_factor_algorithm},
let $c_1\s$ and $c_2\s$, as in \cref{eq:hydro_part_factors}, and $c_3\s = \big(1-c_1\s-c_2\s\big)H_3\s$. The normalized DPFs $c_i'(s) = c_i(s)/\sum_{i=1}^3 c_i(s)$ then gives perfect model matching with $\sum_{i=1}^3 c_i'(s) = 1$.

\subsubsection*{Result}
As seen in the Bode diagram \cref{fig:bode_hydro_wind}, the wind farm compensates for the phase lag of the hydro units at higher frequencies. 
From \cref{prop:NMP_assisted_by_NMP} in the Appendix, we know that that perfect matching can always be achieved, provided that we allow for interaction between the FCR and FFR controllers. For example, in order to achieve $\sum_{i=1}^3c'_i(\jomega) = 1$ for low frequencies where $\Real\big[c_3'(\jomega)\big]<0$, then we need the hydro units to compensate with excessive FCR, i.e., $\Real\big[c_1'(\jomega)+c_2'(\jomega)\big] > 1$. This is seen in \cref{fig:bode_hydro_wind}, where the normalization step increases the gain of the hydro units at $\omega \approx \SI{0.1}{\radian\per\second}$.
A larger separation between the zeros, gives less interaction between the competing NMP dynamics. 
% As a result, the normalization step becomes less significant.

\begin{figure}[tb!]
    \captionsetup[subfloat]{farskip=0pt}
    \centering
    \subfloat[\label{fig:DVPP_wind_lines}Power injections.]
    {{\includegraphics[scale=0.52]{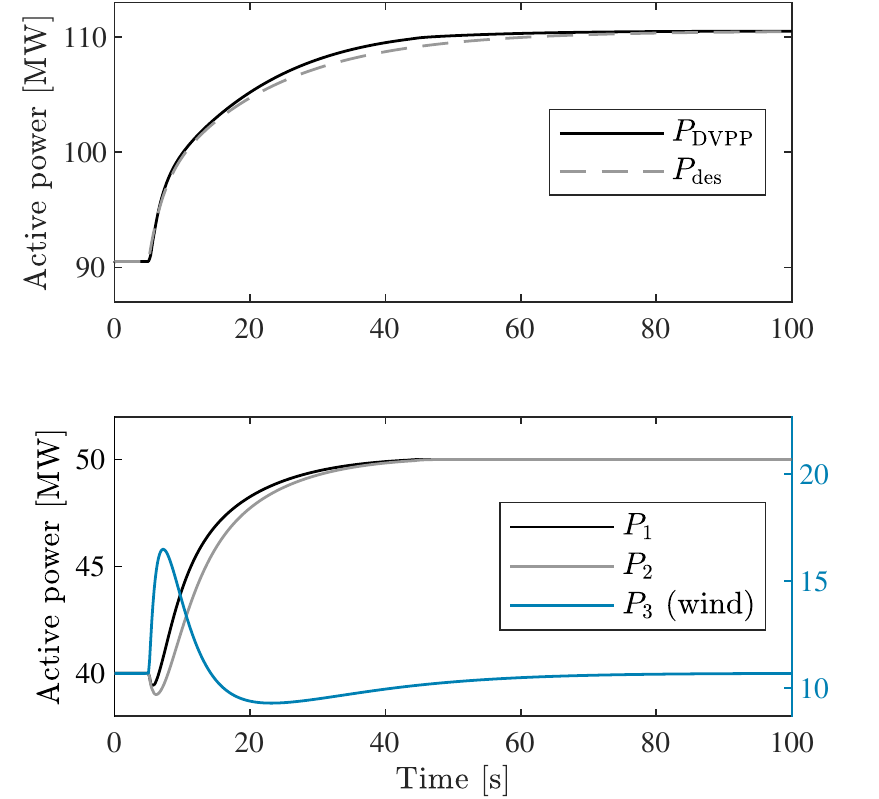}}}
    \hfill
    \subfloat[\label{fig:DVPP_wind}{\WT} response.]
    {{\includegraphics[scale=0.52]{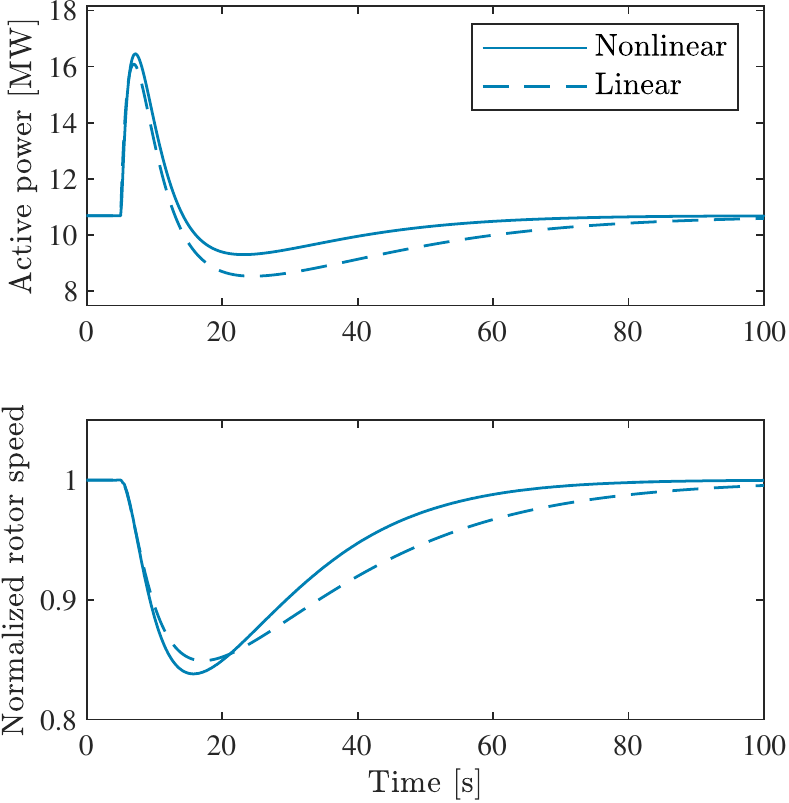}}}
    \caption{Power output of the wind-hydro DVPP.}
    \label{fig:DVPP_wind-hydro}
\end{figure}

The coordinated response to a \SI{1}{\hertz} reference step is shown in \cref{fig:DVPP_wind_lines}. The small discrepancy between the actual DVPP response and the ideal response comes from the fact that the linear model \eqref{eq:H3_wind} underestimates the power output of the nonlinear {\WT} dynamics \cite{bjorkVariablespeedWindTurbineunpublished} as seen in \cref{fig:DVPP_wind}.

\section{Coordinated FCR and FFR in the N5 Test System}
\label{sec:coordinated_in_N5}
In this section, we study FCR and FFR control design in the N5 test system introduced in \cref{sec:Nodic5_intro}.  First we show that, providing FCR solemnly from hydro power, the MP design target \eqref{eq:asmFCRD} cannot be achieved. Then we show that the desired MP design can be achieved by combining hydro and wind.

\subsection{Hydropower FCR in the N5 Test System} 
\begin{figure}[t!]
    \captionsetup[subfloat]{farskip=0pt}
    \centering
    \subfloat[\label{fig:Nordic5_hydro_1} FCR response to a \SI{1400}{\mega\watt} fault.]
    {\makebox[\dimexpr\width+0.3cm\relax][c]{\includegraphics[scale=0.52]{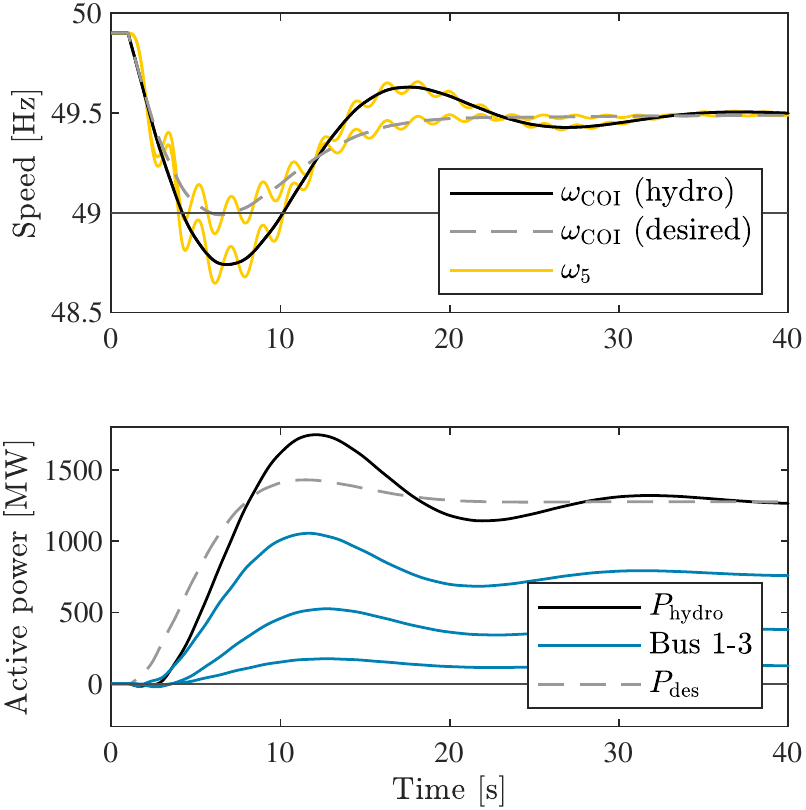}}}
    \hfill
    \subfloat[\label{fig:Nordic5_hydro_bode} Bode diagram.]
    {{\includegraphics[scale=0.52]{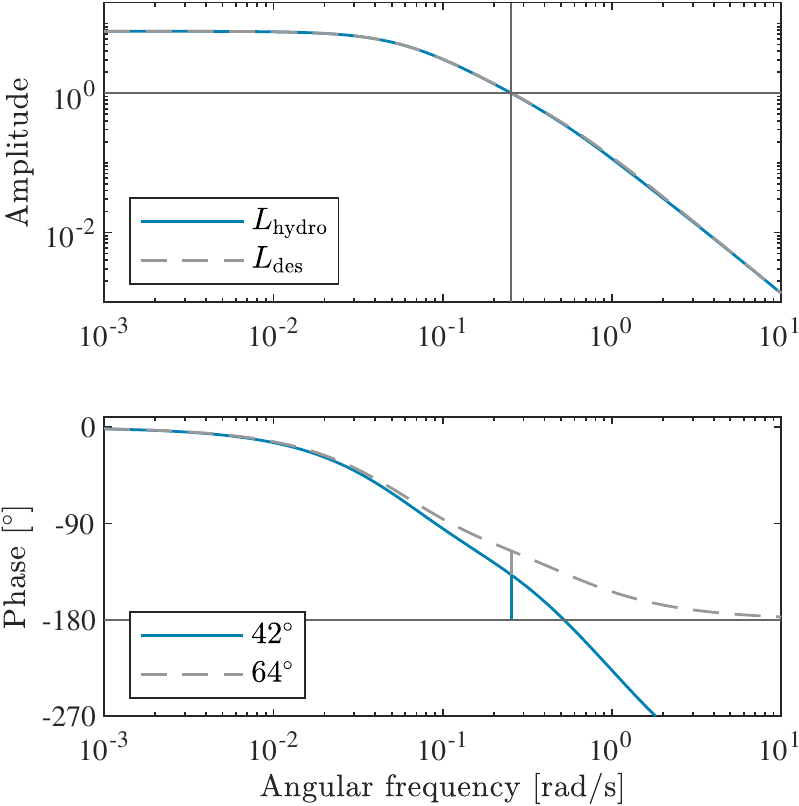}}}
    \caption{N5 test system with hydro FCR.}
    \label{fig:Nordic5_hydro}
\end{figure}

Consider the N5 \SI{110}{\giga\wattsecond} test case studied in \cref{sec:Nodic5_intro}, but now assume that the FCR is provided by hydro units. Based on the test case in \cite{saarinenFullscaleTestModelling2016}, let parameters and FCR resources be distributed according to \cref{tab:Nordic5_hydro}. The FCR controllers for the three hydro units are designed following the same procedure as in \cref{sec:hydropower_FCR}. By design, the target \eqref{eq:asmFCRD} just barely fulfills the FCR-D requirements with ideal actuation and control $P_\mathrm{des} = F_\mathrm{FCR}\s(\omega_\mathrm{ref}-\hat\omega)$. Since the hydro-FCR is NMP, the FCR-D requirements are no longer fulfilled since the combined output at buses 1, 2, and 3, $P_\mathrm{hydro} \neq P_\mathrm{des}$, as seen in \cref{fig:Nordic5_hydro_1}. The reason for this is the negative phase shift, in the aggregated hydro open loop $L_\mathrm{hydro}$ resulting from the NMP zero, shown in \cref{fig:Nordic5_hydro_bode}.

\begin{table}[t!]
\centering
\caption{Hydro parameters for the \SI{110}{\giga\wattsecond} test case.
}
\label{tab:Nordic5_hydro}
\begin{tabular}{cc|ccc}
\hline
  Bus  & FCR [\si{\percent}] & $T_y$ & $T_\mathrm{w}$ & $g_0$ \\
 \hline
 1 & 60 & 0.2 & 0.7 & 0.8 
\\
2 &30 & 0.2 & 1.4 & 0.8 
\\
3 & 10 & 0.2 & 1.4 & 0.8 
\\
\hline
\end{tabular}
\end{table}
%%%%%%%%%%%%

\subsection{Coordinated Wind and Hydropower in the N5 Test System}
\begin{figure}[t!]
    \captionsetup[subfloat]{farskip=0pt}
    \centering
    \subfloat[\label{fig:Nordic5_hydro_wind_1} Response to a \SI{1400}{\mega\watt} fault.]
    {\makebox[\dimexpr\width+0.3cm\relax][c]{\includegraphics[scale=0.52]{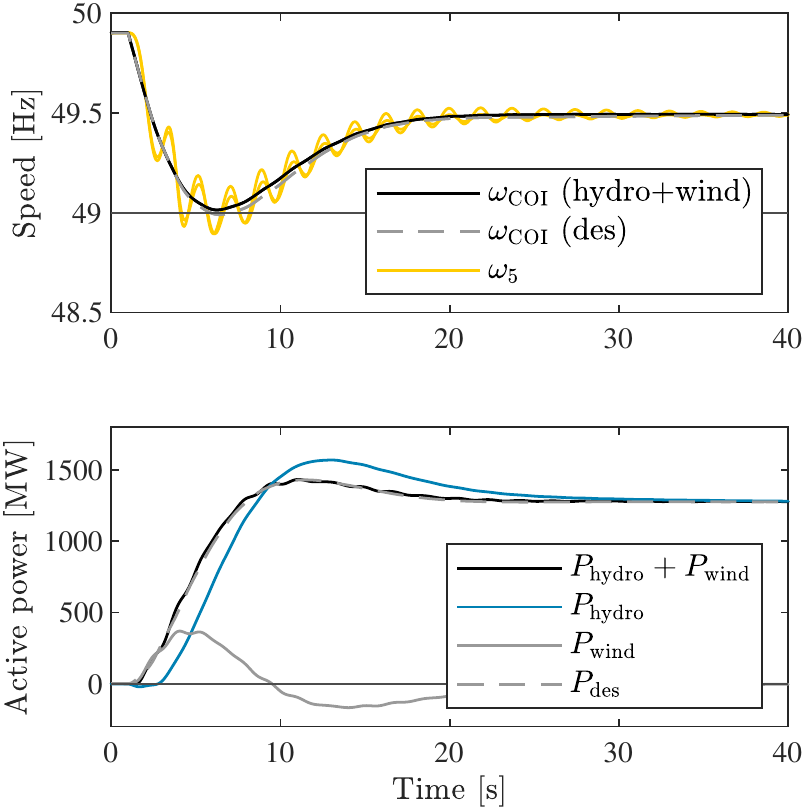}}}
    \hfill
    \subfloat[\label{fig:Nordic5_hydro_wind_bode} Bode diagram.]
    {{\includegraphics[scale=0.52]{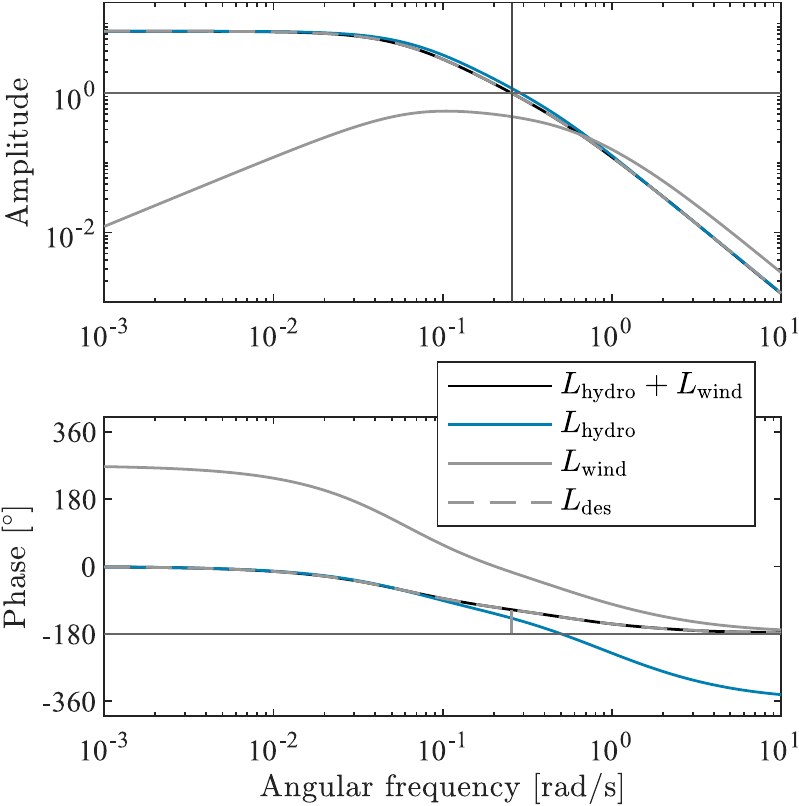}}}
    \caption{N5 test system with coordinated FCR and FFR.}
    \label{fig:Nordic5_hydro_wind}
\end{figure}

Consider again the N5 \SI{110}{\giga\wattsecond} test case, but now let the hydro resources be complemented with FFR from wind power at buses 2 and 4, as shown in \cref{fig:Nordic_5}. Assume that the WTs participating in FFR have a total nominal power rating of \SI{2000}{\mega\watt} distributed according to \cref{tab:Nordic5_wind}. Using the same design procedure described in \cref{sec:wind_power_supported} we leverage the FFR capability of the {\WT}s to achieve perfect matching. With a total output $P_\mathrm{wind}$, the combined wind-hydro FCR and FFR response  closely matches the ideal response $P_\mathrm{des}$, as seen in \cref{fig:Nordic5_hydro_wind_1}. As shown in \cref{fig:Nordic5_hydro_wind_bode}, with the aggregate wind open-loop $L_\mathrm{wind}$, the FCR-D requirements are now fulfilled with no significant change to the cross-over frequency of $L_\mathrm{hydro}$.

The proposed DVPP design method is aligned with the FFR market solution developed to cope with future low-inertia scenarios in the Nordic grid \cite{entso-eFastFrequencyReserve2019}. To address limitations caused by the NMP characteristics of hydro and wind, the proposed DVPP solution targets the whole angular frequency range, so that the interactions between slow and fast dynamics can be addressed when distributing frequency reserves.

\begin{table}[t!]
\centering
\caption{WT parameters for the \SI{110}{\giga\wattsecond} test case. }
\label{tab:Nordic5_wind}
\begin{tabular}{ccc|cc}
\hline
  Bus & $P_\mathrm{nom}$ [\si{\mega\watt}] & $v$  [\si{\meter\per\second}]& $P_\mathrm{MPP}$ [\si{\mega\watt}] & FFR [\si{\percent}] \\
 \hline
2 & \SI{500} & 10 & 348 & 33 
\\
4 & \SI{1500} & 8 & 534 & 67 
\\
\hline
\end{tabular}
\end{table}
%%%%%%%%%%%%

\section{Conclusions}
\label{sec:Conclusion_Ensemble}
A method for distributing ancillary FCR services between a heterogeneous ensemble of controllable plants, by forming a DVPP, has been derived. The method matches the aggregated loop-gain of all participating devices to the Bode diagram of a target function, specified by the SOs requirements. Treating the design as a decentralized model matching problem, the final controller can be implemented with local frequency measurements.
The proposed DVPP design was implemented in a model of the Nordic synchronous grid.
By compensating for the NMP dynamics of hydro, the FCR-D requirements where fulfilled with quite moderate wind resources, without the need for curtailment or battery installations.

There is no automatic control of the tie-line power flows, or area control error (ACE) \cite{kundurPowerSystemStability1994}, in the Nordic grid \cite{saarinenFullscaleTestModelling2016}. This justifies using the COI disturbance response \eqref{eq:coi_dist_simple} for the control design. Instead, ACE is treated indirectly when allocating FCR reserves. Future work will extend the design to incorporate ACE directly. Extending the analysis to include network dynamics also allows for a meaningful representation of voltage dynamics. This allows for including load modeling and voltage control in the control design. We will also extend the
DVPP design, taking device-level constraints into account, uncertainty, and real-time adaption of DPFs.

\appendix
%%%%%%%%%%%%%%%%%%%%%%%%%%%%%%%%%%%%%%%%%%%%%%

%%%%%%%%%%%%%%%%%%%%%%%%%%%%%%%%%%%%%%%%%%%%%%%%%%%%%%%%%%%%%%%%%
%\subsection*{Conditions for Perfect Model Matching with NMP Plants}
\label{app:NMPcond}
\begin{prop}
	\label{prop:sum_is_NMP}%
	Consider two stable plants $c_i(s)$, $i \in\{1,2\}$ with one real NMP zero each at $z_i>0$. Assume positive signed dc gain, i.e., let $c_i(0) > 0$. Then $c(s) = c_1(s)+c_2(s)$ must have at least one real NMP zero $z\in[z_1,z_2]$.
\end{prop}
\begin{IEEEproof}%
% Let  $	c_i(s) = a_i(s) ({z_i-s})/b(s)$, where $b\s$ and $a_i\s$ are polynomials with no RHP roots.
Let  $c_i(s) = a_i(s) ({z_i-s})/b_i(s)$, where $b_i\s$ and $a_i\s$ are polynomials with no RHP roots. Without loss of generality, assume that plants have been normalized so that $b_1\s=b_2\s=b\s$.
Then $c\s=c_1\s+c_2\s$ is
	\begin{equation}
	c(s)  = 
	\Big({z_1a_1(s) + z_2a_2(s) - s\big(a_1(s) + a_2(s)\big)}\Big)\Big/{b(s)} ,
	\end{equation}
and has zeros on the positive real axis where
	\begin{equation}
	\label{eq:apx1}
	    \frac{z_1a_1(\sigma) + z_2a_2(\sigma)}{a_1(\sigma)+a_2(\sigma)} - \sigma = 0, \quad \sigma \geq 0.
	\end{equation}
    With $c_i(0) > 0$, and with no zeros in the RHP, $a_i(\sigma)>0$, $\forall \sigma \geq 0$.
The first term in \eqref{eq:apx1} is therefore a convex combination
	\begin{equation}
	     \frac{z_1 a_1(\sigma) + z_2 a_2(\sigma) }{a_1(\sigma) + a_2(\sigma)} \in [z_1,z_2], \quad \forall \sigma \geq 0.
	\end{equation}
Since $0< \min(z_1,z_2) \leq \max(z_1,z_2)<\infty$,
\eqref{eq:apx1} must have at least one real NMP zero where $\sigma = z\in[z_1,z_2]$.
\end{IEEEproof}

\begin{prop}
	\label{prop:NMP_assisted_by_MP}%
	Consider a stable and proper plant $c_1(s)$. If $\Real \big[c_1(\jomega) \big] \leq 1, \ \forall \omega$, then perfect matching is achieved with the stable and MP plant $c_2(s)= 1-c_1(s)$.
\end{prop}
\begin{IEEEproof}%
	If $c_1(s)$ is stable and proper, then so is $c_2(s) = 1-c_1(s)$. If $\Real \big[c_1(\jomega)\big]\leq 1$, $\forall \omega$, then $\Real \big[ c_2(\jomega) \big]\geq 0$, $\forall \omega$. 
	Since $c_2(s)$ is stable and positive real, it is also MP \cite{ioannouFrequencyDomainConditions1987}.
% 	Since $c_2(s)$ has to be proper, and since the $n_p$ stable poles give $\SI{90}{\degree}$ phase lag each, $c_2(s)$ needs to have $n_p$ or $n_p-1$ closed left half plane zeros and no RHP zeros.
\end{IEEEproof}

\begin{prop}
	\label{prop:NMP_assisted_by_NMP}%
	Consider a stable first-order plant
	\begin{equation}
	    c_1(s) = (1+\varepsilon) \frac{z_1-s}{s+z_1},
	\end{equation}
	with RHP zero $z_1>0$ and $\varepsilon>0$ so that $c_1(0) >1$. Then under perfect matching, the assisting first-order plant $c_2(s) = 1-c_1(s)$ is stable, has a RHP zero in
% 	\begin{equation}
$
	    z_2 = z_1{\varepsilon}/({2+\varepsilon}),
	    $
% 	\end{equation}
	and has a negative steady-gain $c_2(0) = -\varepsilon < 0$.
\end{prop}
\begin{IEEEproof}% 
The assisting plant $c_2(s)$ is
	\begin{equation}
1-c_1(s)=  \frac{s(2+\varepsilon) - z_1 \varepsilon   }{s+z_1} 
	=
	\frac{2+\varepsilon}{s+z_1}\left({s-\frac{z_1\varepsilon}{2+\varepsilon}}\right).
	\null \hfill \IEEEQEDhereeqn
	\end{equation}
\end{IEEEproof}

%%%%%%%%%%%%%%%%%%%%%%%%%%%%%%%%%%%%%%%%%%%%%%%%%%%%%%%%%%%%

\bibliographystyle{IEEEtran}

\bibliography{bib_ensemble.bib,bib_online.bib}

\begin{thebibliography}{10}
\providecommand{\url}[1]{#1}
\csname url@samestyle\endcsname
\providecommand{\newblock}{\relax}
\providecommand{\bibinfo}[2]{#2}
\providecommand{\BIBentrySTDinterwordspacing}{\spaceskip=0pt\relax}
\providecommand{\BIBentryALTinterwordstretchfactor}{4}
\providecommand{\BIBentryALTinterwordspacing}{\spaceskip=\fontdimen2\font plus
\BIBentryALTinterwordstretchfactor\fontdimen3\font minus
  \fontdimen4\font\relax}
\providecommand{\BIBforeignlanguage}[2]{{%
\expandafter\ifx\csname l@#1\endcsname\relax
\typeout{** WARNING: IEEEtran.bst: No hyphenation pattern has been}%
\typeout{** loaded for the language `#1'. Using the pattern for}%
\typeout{** the default language instead.}%
\else
\language=\csname l@#1\endcsname
\fi
#2}}
\providecommand{\BIBdecl}{\relax}
\BIBdecl

\bibitem{dall'aneseOptimalRegulationVirtual2018}
E.~Dall'Anese, S.~S. Guggilam, A.~Simonetto, Y.~C. Chen, and S.~V. Dhople,
  ``Optimal regulation of virtual power plants,'' \emph{IEEE Trans. Power
  Syst.}, vol.~33, no.~2, pp. 1868--1881, Mar. 2018.

\bibitem{milanoFoundationsChallengesLowinertia2018}
F.~Milano, F.~D{\"o}rfler, G.~Hug, D.~J. Hill, and G.~Verbi{\v c},
  ``Foundations and challenges of low-inertia systems (invited paper),'' in
  \emph{Power {{Systems Computation Conference}}}, {Dublin, Ireland}, Jun.
  2018.

\bibitem{rahmanLargestBlackoutsWorld2016}
K.~M.~J. Rahman, M.~M. Munnee, and S.~Khan, ``Largest blackouts around the
  world: {{Trends}} and data analyses,'' in \emph{{{IEEE International WIE
  Conference}} on {{Electrical}} and {{Computer Engineering}}}, {Pune, India},
  Dec. 2016, pp. 155--159.

\bibitem{brundlingerReviewAssessmentLatest2016}
R.~Br{\"u}ndlinger, ``Review and assessment of latest grid code developments in
  {{Europe}} and selected international markets with respect to high
  penetration {{PV}},'' in \emph{6th {{Solar Integration Workshop}}}, {Vienna,
  Austria}, Nov. 2016.

\bibitem{sabooriVirtualPowerPlant2011}
H.~Saboori, M.~Mohammadi, and R.~Taghe, ``Virtual power plant ({{VPP}}),
  definition, concept, components and types,'' in \emph{Asia-{{Pacific Power}}
  and {{Energy Engineering Conference}}}, {Wuhan, China}, Mar. 2011.

\bibitem{ghavidelReviewVirtualPower2016}
S.~Ghavidel, L.~Li, J.~Aghaei, T.~Yu, and J.~Zhu, ``A review on the virtual
  power plant: {{Components}} and operation systems,'' in \emph{{{IEEE
  International Conference}} on {{Power System Technology}}}, {Wollongong,
  Australia}, Sep. 2016.

\bibitem{vasiraniAgentBasedApproachVirtual2013}
M.~Vasirani, R.~Kota, R.~L.~G. Cavalcante, S.~Ossowski, and N.~R. Jennings,
  ``An agent-based approach to virtual power plants of wind power generators
  and electric vehicles,'' \emph{IEEE Trans. Smart Grid}, vol.~4, no.~3, pp.
  1314--1322, Sep. 2013.

\bibitem{alvarezGenericStorageModel2019}
M.~Alvarez, S.~K. R{\"o}nnberg, J.~Berm{\'u}dez, J.~Zhong, and M.~H.~J. Bollen,
  ``A generic storage model based on a future cost piecewise-linear
  approximation,'' \emph{IEEE Trans. Smart Grid}, vol.~10, no.~1, pp. 878--888,
  Jan. 2019.

\bibitem{moutisVoltageRegulationSupport2018}
P.~Moutis, P.~S. Georgilakis, and N.~D. Hatziargyriou, ``Voltage regulation
  support along a distribution line by a virtual power plant based on a center
  of mass load modeling,'' \emph{IEEE Trans. Smart Grid}, vol.~9, no.~4, pp.
  3029--3038, Jul. 2018.

\bibitem{alhelouPrimaryFrequencyResponse2020}
H.~H. Alhelou, P.~Siano, M.~Tipaldi, R.~Iervolino, and F.~Mahfoud,
  ``\BIBforeignlanguage{en}{Primary frequency response improvement in
  interconnected power systems using electric vehicle virtual power plants},''
  \emph{\BIBforeignlanguage{en}{World Electric Vehicle Journal}}, vol.~11,
  no.~2, p.~40, May 2020.

\bibitem{schifferSynchronizationDroopcontrolledMicrogrids2013}
J.~Schiffer, D.~Goldin, J.~Raisch, and T.~Sezi, ``Synchronization of
  droop-controlled microgrids with distributed rotational and electronic
  generation,'' in \emph{52nd {{IEEE Conference}} on {{Decision}} and
  {{Control}}}, {Firenze, Italy}, Dec. 2013, pp. 2334--2339.

\bibitem{zhongCoordinatedControlVirtual2021}
W.~Zhong, J.~Chen, M.~Liu, M.~A.~A. Murad, and F.~Milano,
  ``\BIBforeignlanguage{en}{Coordinated control of virtual power plants to
  improve power system short-term dynamics},''
  \emph{\BIBforeignlanguage{en}{Energies}}, vol.~14, no.~4, p. 1182, Feb. 2021.

\bibitem{posytyfConceptObjectives2021_manual}
\BIBentryALTinterwordspacing
POSYTYF. Concept \& objectives. (2021, Feb 16). [Online]. Available:
  \url{https://posytyf-h2020.eu/project-overview/work-plan}
\BIBentrySTDinterwordspacing

\bibitem{melkiInvestigationFrequencyContainment2019}
J.~Melki and H.~Ghasemi, ``\BIBforeignlanguage{en}{Investigation of frequency
  containment reserves with inertial response and batteries},'' Bachelor's
  Thesis, KTH Royal Institute of Technology, {Stockholm}, 2019.

\bibitem{liPowerElectronicInterface2008}
W.~Li and G.~Joos, ``A power electronic interface for a battery supercapacitor
  hybrid energy storage system for wind applications,'' in \emph{{{IEEE Power
  Electronics Specialists Conference}}}, {Rhodes, Greece}, Jun. 2008, pp.
  1762--1768.

\bibitem{glavinStandalonePhotovoltaicSupercapacitor2008}
M.~E. Glavin, P.~K.~W. Chan, S.~Armstrong, and W.~G. Hurley, ``A stand-alone
  photovoltaic supercapacitor battery hybrid energy storage system,'' in
  \emph{13th {{International Power Electronics}} and {{Motion Control
  Conference}}}, {Poznan, Poland}, Sep. 2008, pp. 1688--1695.

\bibitem{mullerLargescaleDemonstrationPrecise2019}
F.~L. M{\"u}ller and B.~Jansen, ``\BIBforeignlanguage{en}{Large-scale
  demonstration of precise demand response provided by residential heat
  pumps},'' \emph{\BIBforeignlanguage{en}{Applied Energy}}, vol. 239, pp.
  836--845, Apr. 2019.

\bibitem{morrenWindTurbinesEmulating2006}
J.~Morren, S.~W.~H. de~Haan, W.~L. Kling, and J.~A. Ferreira, ``Wind turbines
  emulating inertia and supporting primary frequency control,'' \emph{IEEE
  Trans. Power Syst.}, vol.~21, no.~1, pp. 433--434, Feb. 2006.

\bibitem{zhaoFastFrequencySupport2020}
X.~Zhao, Y.~Xue, and X.-P. Zhang, ``Fast frequency support from wind turbine
  systems by arresting frequency nadir close to settling frequency,''
  \emph{IEEE Open Access Journal of Power and Energy}, vol.~7, pp. 191--202,
  2020.

\bibitem{wilches-bernalFundamentalStudyApplying2016}
F.~{Wilches-Bernal}, J.~H. Chow, and J.~J. {Sanchez-Gasca}, ``A fundamental
  study of applying wind turbines for power system frequency control,''
  \emph{IEEE Trans. Power Syst.}, vol.~31, no.~2, pp. 1496--1505, Mar. 2016.

\bibitem{agneholmFCRDDesignRequirements2019}
E.~Agneholm, S.~A. Meybodi, M.~Kuivaniemi, P.~Ruokolainen, J.~N.
  {\O}deg{\aa}rd, N.~Modig, and R.~Eriksson,
  ``\BIBforeignlanguage{en}{{{FCR}}-{{D}} design of requirements \textendash{}
  phase 2},'' {ENTSOE-E}, Tech. Rep., 2019.

\bibitem{entso-eFastFrequencyReserve2019}
{ENTSO-E}, ``\BIBforeignlanguage{en}{Fast frequency reserve \textendash{}
  solution to the {{Nordic}} inertia challenge},'' Tech. Rep., 2019.

\bibitem{kundurPowerSystemStability1994}
P.~Kundur, \emph{\BIBforeignlanguage{eng}{Power {{System Stability}} and
  {{Control}}}}.\hskip 1em plus 0.5em minus 0.4em\relax {New York}:
  {McGraw-Hill}, 1994.

\bibitem{saarinenFullscaleTestModelling2016}
L.~Saarinen, P.~Norrlund, U.~Lundin, E.~Agneholm, and A.~Westberg, ``Full-scale
  test and modelling of the frequency control dynamics of the {{Nordic}} power
  system,'' in \emph{{{IEEE Power}} and {{Energy Society General Meeting}}},
  {Boston, MA}, Jul. 2016.

\bibitem{hydro-quebecSimscapeElectricalReference2020_manual}
{Hydro-Qu{\'e}bec} and {The MathWorks}, ``\BIBforeignlanguage{en}{Simscape
  {{Electrical Reference}} ({{Specialized Power Systems}})},'' {Natick, MA},
  Tech. Rep., 2020.

\bibitem{entso-eNordicSynchronousArea2018}
{ENTSO-E}, ``\BIBforeignlanguage{en}{Nordic synchronous area proposal for
  additional properties of {{FCR}} in accordance with {{Article}} 154(2) of the
  {{Commission Regulation}} ({{EU}}) 2017/1485 of 2 {{August}} 2017
  establishing a guideline on electricity transmission system operation},''
  2018.

\bibitem{NordLink_manual}
\BIBentryALTinterwordspacing
{Statnett}. {NordLink}. (2021, Feb 13). [Online]. Available:
  \url{https://www.statnett.no/en/our-projects/interconnectors/nordlink/}
\BIBentrySTDinterwordspacing

\bibitem{jonkmanDefinition5MWReference2009}
J.~Jonkman, S.~Butterfield, W.~Musial, and G.~Scott,
  ``\BIBforeignlanguage{en}{Definition of a 5-{{MW}} reference wind turbine for
  offshore system development},'' {NREL}, {USA}, Tech. Rep., 2009.

\bibitem{bjorkVariablespeedWindTurbineunpublished}
J.~Bj{\"o}rk, D.~V. Pombo, and K.~H. Johansson, ``Variable-speed wind turbine
  control designed for coordinated fast frequency reserves,'' unpublished.

\bibitem{skogestadMultivariableFeedbackControl2007}
S.~Skogestad and I.~Postlethwaite, \emph{Multivariable {{Feedback Control}}:
  {{Analysis}} and {{Design}}}, 2nd~ed.\hskip 1em plus 0.5em minus 0.4em\relax
  {New York}: {Wiley}, 2007.

\bibitem{zhouRobustOptimalControl1996}
K.~Zhou, \emph{\BIBforeignlanguage{eng}{Robust and {{Optimal Control}}}}.\hskip
  1em plus 0.5em minus 0.4em\relax {Englewood Cliffs, NJ}: {Prentice Hall},
  1996.

\bibitem{patesRobustScaleFreeSynthesis2019}
R.~Pates and E.~Mallada, ``Robust scale-free synthesis for frequency control in
  power systems,'' \emph{IEEE Control Netw. Syst.}, vol.~6, no.~3, pp.
  1174--1184, Sep. 2019.

\bibitem{ioannouFrequencyDomainConditions1987}
P.~Ioannou and G.~Tao, ``Frequency domain conditions for strictly positive real
  functions,'' \emph{IEEE Trans. Autom. Control}, vol.~32, no.~1, pp. 53--54,
  Jan. 1987.

\end{thebibliography}

\end{document}